\documentclass[aps,prl,floatfix,twocolumn,reprint,amsmath,amssymb,superscriptaddress]{revtex4-1}

\usepackage{bm}
\usepackage{times}
\usepackage{graphicx}
\usepackage{color}
\usepackage{dcolumn}
\usepackage[colorlinks=true, letterpaper=true, pdfstartview=FitV, linkcolor=blue, citecolor=blue, urlcolor=blue]{hyperref}
\usepackage{appendix}
\usepackage[normalem]{ulem}
\usepackage{soul}
\usepackage{makecell} 
\usepackage[figuresright]{rotating} 

\begin{document}
	
	\title{Crystal Thermal Transport in Altermagnetic RuO$_2$}
	
	\author{Xiaodong Zhou}
	\affiliation{Centre for Quantum Physics, Key Laboratory of Advanced Optoelectronic Quantum Architecture and Measurement (MOE),School of Physics, Beijing Institute of Technology, Beijing 100081, China}
	\affiliation{Beijing Key Lab of Nanophotonics and Ultrafine Optoelectronic Systems, School of Physics, Beijing Institute of Technology, Beijing 100081, China}    
	\affiliation {Laboratory of Quantum Functional Materials Design and Application, School of Physics and Electronic Engineering, Jiangsu Normal University, Xuzhou 221116, China}
	
	\author{Wanxiang Feng}
	\email{wxfeng@bit.edu.cn}
	\affiliation{Centre for Quantum Physics, Key Laboratory of Advanced Optoelectronic Quantum Architecture and Measurement (MOE),School of Physics, Beijing Institute of Technology, Beijing 100081, China}
	\affiliation{Beijing Key Lab of Nanophotonics and Ultrafine Optoelectronic Systems, School of Physics, Beijing Institute of Technology, Beijing 100081, China}    
	
	\author{Run-Wu Zhang}
	\affiliation{Centre for Quantum Physics, Key Laboratory of Advanced Optoelectronic Quantum Architecture and Measurement (MOE),School of Physics, Beijing Institute of Technology, Beijing 100081, China}
	\affiliation{Beijing Key Lab of Nanophotonics and Ultrafine Optoelectronic Systems, School of Physics, Beijing Institute of Technology, Beijing 100081, China}    
	
	\author{Libor {\v S}mejkal}
	\affiliation{Institute of Physics, Johannes Gutenberg University Mainz, 55099 Mainz, Germany}
	\affiliation{Institute of Physics, Czech Academy of Sciences, Cukrovarnick{\'a} 10, 162 00 Praha 6, Czech Republic}
	
	\author{Jairo Sinova}
	\affiliation{Institute of Physics, Johannes Gutenberg University Mainz, 55099 Mainz, Germany}
	\affiliation{Institute of Physics, Czech Academy of Sciences, Cukrovarnick{\'a} 10, 162 00 Praha 6, Czech Republic}
	
	\author{Yuriy Mokrousov}
	\affiliation{Institute of Physics, Johannes Gutenberg University Mainz, 55099 Mainz, Germany}
	\affiliation{Peter Gr\"unberg Institut and Institute for Advanced Simulation, Forschungszentrum J\"ulich and JARA, 52425 J\"ulich, Germany}
	
	\author{Yugui Yao}
	\email{ygyao@bit.edu.cn}
	\affiliation{Centre for Quantum Physics, Key Laboratory of Advanced Optoelectronic Quantum Architecture and Measurement (MOE),School of Physics, Beijing Institute of Technology, Beijing 100081, China}
	\affiliation{Beijing Key Lab of Nanophotonics and Ultrafine Optoelectronic Systems, School of Physics, Beijing Institute of Technology, Beijing 100081, China}    
	
	\date{\today}
	
	\begin{abstract}
		
		We demonstrate the emergence of a pronounced thermal transport in the recently discovered class of magnetic materials $-$ altermagents. From symmetry arguments and first principles calculations performed for the showcase altermagnet, RuO$_2$, we uncover that crystal Nernst and crystal thermal Hall effects in this material are very large and strongly anisotropic with respect to the N\'eel vector.  We find the large crystal thermal transport to originate from three sources of Berry's curvature in momentum space: the Weyl fermions due to crossings between well-separated bands, the strong spin-flip pseudo-nodal surfaces, and the weak spin-flip ladder transitions, defined by transitions among very weakly spin-split states of similar dispersion crossing the Fermi surface.  Moreover, we reveal that the anomalous thermal and electrical transport coefficients in RuO$_2$ are linked by an extended Wiedemann-Franz law in a temperature range much wider than expected for conventional magnets. Our results suggest that altermagnets may assume a leading role in realizing concepts in spincaloritronics not achievable with ferromagnets or antiferromagnets.
		
	\end{abstract}
	
	\maketitle
	
\begin{figure}
		\centering
		\includegraphics[width=\columnwidth]{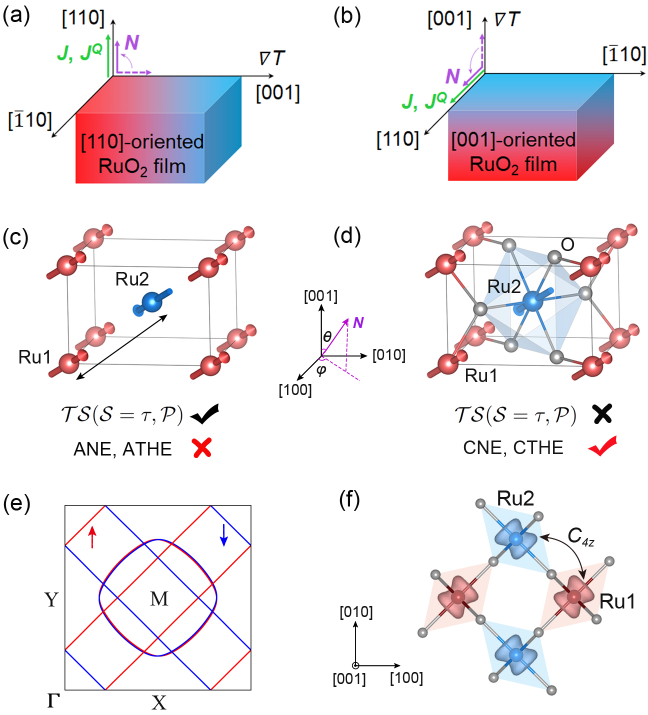}
		\caption{(a,b) Schematics of anomalous Nernst and anomalous thermal Hall effects in [110] and [001] oriented RuO$_2$ films.  $\textbf{\textit{N}}$ denotes the N{\'e}el vector rotating from the [001] to [110] axis on the ($\bar{1}10$) plane.  $J$ and $J^{Q}$ are detectable electrical and thermal currents, respectively, induced by a temperature gradient $\nabla T$.  (c,d) Magnetic unit cell of RuO$_2$ without and with O atoms.  Red and blue balls represent two Ru atoms with antiparallel spin magnetic moments, and gray balls represent nonmagnetic O atoms.  The anomalous thermal transport is prohibited by the $\mathcal{TS}$-symmetry of magnetic lattice alone, however, this symmetry is broken when taking into account the cage of O atoms, which gives rise to ``crystal" thermal transport.  Inset denotes the parametrization of the N{\'e}el vector \textbf{\textit{N}}($\varphi, \theta$) in spherical coordinates.  (e) Reciprocal space Fermi surface featured by strong (cigar shaped) and weak (circular) spin-splitting states.  (f) Real space alternating spin density related by four-fold crystal rotation.}
		\label{fig1}
	\end{figure}

	The anomalous Nernst effect (ANE)~\cite{D-Xiao2006} and anomalous thermal Hall effect (ATHE) (or anomalous Righi-Leduc effect)~\cite{T-Qin2011}---the thermoelectric and thermal analogues of the anomalous Hall effect (AHE)---are the two fundamental anomalous thermal transport manifestations in spin caloritronics~\cite{Bauer2012,Boona2014}. They describe separately the appearance of a transverse electrical current density $J_j$ and a transverse thermal current density $J_j^Q$ induced by a longitudinal temperature gradient $-\nabla_i T$ in the absence of an external magnetic field [Figs.~\ref{fig1}(a) and~\ref{fig1}(b)]. The ANE and ATHE are usually expressed as~\cite{Callen1948}
	\begin{eqnarray}
		\textit{J}_j  &=& \alpha_{ji}(-\nabla_i T), \\
		\textit{J}_j^Q &=& \kappa_{ji}(-\nabla_i T),
	\end{eqnarray}
	where $\alpha_{ji}$ is the anomalous Nernst conductivity (ANC) and $\kappa_{ji}$ is the anomalous thermal Hall conductivity (ATHC). The transverse thermal and electrical transport are related by the  Lorenz ratio, $L_{ij} = \kappa_{ij}/(\sigma_{ij} T)$, where $\sigma_{ij}$ is the anomalous Hall conductivity (AHC).  In zero temperature limit, the Wiedemann-Franz (WF) law, $L_{ij} (T\rightarrow0) = \pi^2k_B^2/3e^2 = L_0$ (Sommerfeld constant), has been witnessed in conventional ferromagnets~\cite{Onose2008,Shiomi2009,Shiomi2010}.
	
	The ANE and ATHE have been investigated extensively in ferromagnets, and they are generally proportional to the spontaneous magnetization~\cite{Smith1911,Smith1921,Lee2004,Miyasato2007,Onoda2008,Onose2008, Shiomi2009,Shiomi2010}. The antiferromagnets (AFMs) are long presumed to have vanishing anomalous thermal transport due to vanishing macroscopic magnetization.  However, this assumption has been challenged in certain noncollinear AFMs hosting nonvanishing flavors of spin chirality. For example, the ANE and ATHE are predicted theoretically and/or observed experimentally in coplanar noncollinear AFMs Mn$_{3}Y$ ($Y$ = Ge, Sn)~\cite{Ikhlas2017,XK-Li2017,GY-Guo2017,LC-Xu2020,Sugii2019} and Mn$_{3}X$N ($X$ = Ga, Zn, Ag, Ni)~\cite{XD-Zhou2020,XD-Zhou2019a}, where the vector spin chirality plays a crucial role~\cite{XD-Zhou2019a}. The ANE and ATHE are also found in noncoplanar AFMs or skyrmions~\cite{Shiomi2013,Hirschberger2020,H-Zhang2021,Owerre2017,YL-Lu2019}, in which the scalar spin chirality generates a real-space Berry phase of propagating electrons. The ANE and ATHE can emerge in noncoplanar magnets even without spin-orbit coupling (SOC), like the (quantum) topological Hall and topological magneto-optical effects~\cite{J-Zhou2016,Hanke2017,WX-Feng2020}.
	
	While the ANE and ATHE in noncollinear AFMs were explored in recent past, much less is known about their properties in collinear AFMs. Conventional collinear AFMs host time-reversal-like degenerate electronic bands  through a combined $\mathcal{TS}$ symmetry ($\mathcal{S}$ being spatial inversion $\mathcal{P}$ or translation $\tau$, $\mathcal{T}$ being time-reversal), prohibiting ANE and ATHE [Fig.~\ref{fig1}(c)]. The $\mathcal{TS}$ symmetry breaking by nonmagnetic atoms can have a drastic effect, as it e.g. activates the ``crystal" Hall effect in  RuO$_2$~\cite{Smejkal2020,ZX-Feng2022} by the virtue of  crystal time-reversal symmetry breaking. Respectively, from rigorous spin symmetry analysis, RuO$_2$ is recognized as a representative of a distinct flavour of magnetism beyond ferromagnetism and antiferromagnetism, termed altermagnetism~\cite{Smejkal2022b,Smejkal2022a}. Altermagnets are characterized by spin polarization in both reciprocal and real spaces~\cite{Smejkal2022b,Smejkal2022a,HY-Ma2021} [Figs.~\ref{fig1}(e) and~\ref{fig1}(f)]. Thus, the question arises whether analogous ``crystal" thermal effects [Fig.~\ref{fig1}(d)] emerge in altermagnets due to a non-trivial interplay of spin and crystal symmetries.

	Here, based on symmetry analysis and state-of-the-art first-principles calculations, we answer this question positively. Taking RuO$_2$ as a showcase representative of altermagnetism, we reveal a novel class of anomalous thermal transport effects, namely the crystal Nernst effect (CNE) and crystal thermal Hall effect (CTHE), to emerge in altermagnetic materials.  The crystal thermal transport in RuO$_2$ is strongly anisotropic with respect to the orientation of the N{\'e}el vector, and can be  observed experimentally in [110] or [001] oriented films.  We elucidate that large crystal thermal transport is microscopically related to Weyl fermions, pseudo-nodal surfaces, and ladder transitions.  Moreover, we find that the anomalous Lorenz ratio is in accordance with the Sommerfeld constant in a rather wide temperature range of 0--150 K.  Our results suggest a promising altermagnetic spin-caloritronics platform for the realization of room-temperature crystal thermal transport.
	
	The material under study, RuO$_2$, has a rutile structure with a space group $P4_2/mnm$ [Fig.~\ref{fig1}(d)]. Several recent studies reveal room-temperature collinear antiparallel magnetic order in both RuO$_2$ thin films and bulk crystals~\cite{Berlijn2017,ZH-Zhu2019,ZX-Feng2022,Lovesey2022}, which promotes exciting applications in spintronics, such as the crystal Hall effect~\cite{Smejkal2020,ZX-Feng2022}, spin polarized currents\cite{Gonzalez2021,Smejkal2022}, spin-splitter torque~\cite{Gonzalez2021,Bose2022,Karube2022,H-Bai2022}, giant tunneling magnetoresistance~\cite{DF-Shao2021a,Smejkal2022}, magneto-optical effect~\cite{Samanta2020,XD-Zhou2021}, and tunneling AHE~\cite{DF-Shao2021a,DF-Shao2022}. %However, its most promising feature---metallic collinear magnetism up to the N{\'e}el temperature of 300--400 K~\cite{ZH-Zhu2019}---has not been exploited in full yet.
	
	The N{\'e}el vector of RuO$_2$, $\textbf{\textit{N}}=\textbf{\textit{S}}_{1}-\textbf{\textit{S}}_{2}$ ($\textbf{\textit{S}}_{1,2}$ are antiparallel spin moments on two Ru atoms), was shown to exhibit a dominant projection along the [001] axis~\cite{Berlijn2017,ZH-Zhu2019}.  For convenience, we introduce the spherical coordinates to describe the orientation of the N{\'e}el vector  \textbf{\textit{N}}($\varphi, \theta$) (see Fig.~\ref{fig1}), with the experimentally observed direction of  \textbf{\textit{N}}($\varphi=107.2^\circ, \theta=22.7^\circ$)~\cite{ZH-Zhu2019}.  Such a low-symmetry orientation is very susceptible to experimental conditions, and it can be changed by an external magnetic field or by the effects of spin-orbit torque~\cite{Wadley2016,Godinho2018,XZ-Chen2019}.  Notably, a recent experiment has realized the crystal AHE in RuO$_2$ films utilizing the magnetic field-tunable reorientation of the spin-quantization axis~\cite{ZX-Feng2022}. Here we focus on the crystal thermal transport in RuO$_2$ assuming that the N{\'e}el vector can rotate from the [001] to [110] axis on the ($\bar{1}10$) plane, i.e., \textbf{\textit{N}}($ \varphi = 45^\circ, 0^\circ \leq \theta \leq 360^\circ $).
	
	\begin{figure*}
		\centering
		\includegraphics[width=2\columnwidth]{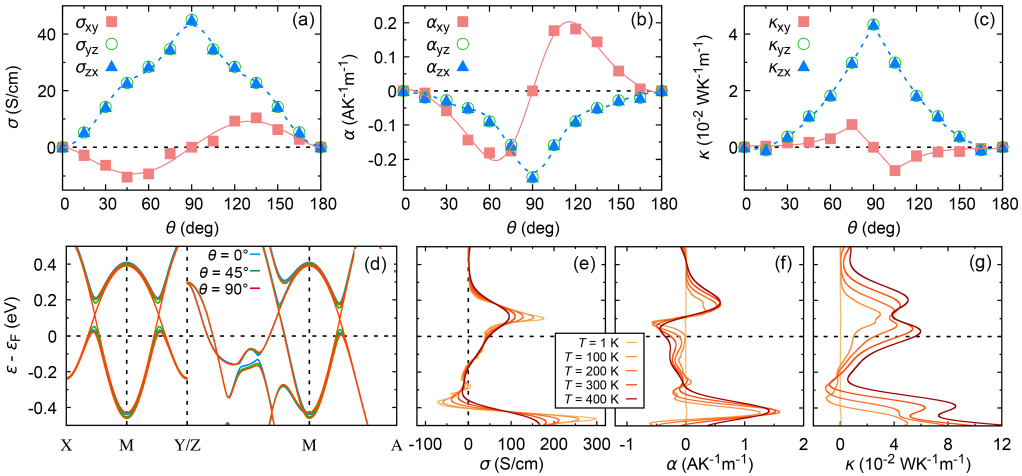}
		\caption{(a-c) Components of the anomalous Hall conductivity $\sigma$, anomalous Nernst conductivity $\alpha$, and anomalous thermal Hall conductivity $\kappa$ at $T=$ 300 K as a function of polar angle $\theta$ when the N{\'e}el vector rotates within the ($\bar{1}10$) plane. The lines are guides to the eye. The conductivity at $\theta + \pi$ is not plotted as it is opposite in sign to that at $\theta$ due to time-reversal symmetry.  (d) Relativistic band structures of RuO$_2$ for \textbf{\textit{N}}($\varphi=45^\circ, \theta=0^\circ, 45^\circ, 90^\circ$).  (e-g) $\sigma$, $\alpha$, and $\kappa$ as a function of Fermi energy at \textbf{\textit{N}}($\varphi=45^\circ, \theta=90^\circ$) for different temperatures. The color scheme from light to dark corresponds to ``warming" from 1 K to 400 K.}
		\label{fig2}
	\end{figure*}
	
	We assess the anomalous transport coefficients as
	%within the Landauer-B{\"u}ttiker formalism
	~\cite{Ashcroft1976,Houten1992,Behnia2015}
	\begin{equation}
		R_{ij}^{(n)} = \int_{-\infty}^\infty(\varepsilon-\mu)^n (-\dfrac{\partial f}{\partial \varepsilon}) \sigma_{ij}^{T = 0}(\varepsilon) d\varepsilon, \label{eq:R}
	\end{equation}
	where $\mu$ is chemical potential, $f =1/[\text{exp}((\varepsilon-\mu)/k_{B}T)+1]$ is the Fermi-Dirac distribution function, and $\sigma_{ij}^{T = 0}(\varepsilon)$ is zero-temperature intrinsic AHC (see Supplemental Material~\cite{SuppMater}). Then, the temperature-dependent AHC $\sigma$, ANC $\alpha$, and ATHC $\kappa$ respectively read
	\begin{equation}
		\sigma_{ij} = R_{ij}^{(0)}, \quad
		\alpha_{ij} = -R_{ij}^{(1)}/eT, \quad
		\kappa_{ij} = R_{ij}^{(2)}/e^2T. \label{eq:kappa}
	\end{equation}
	The computed three crystal thermal transport coefficients of RuO$_2$, shown in Figs.~\ref{fig2}(a)-(c), are remarkably all nonzero except for \textbf{\textit{N}}($\varphi = 45^{\circ}, \theta = \mathbb{N}\pi$).  For the orientation of the N{\'e}el vector along the [001] axis ($\theta = \mathbb{N}\pi$), the effects are prohibited by the presence of multiple mirror planes: the glide mirror $\mathcal{M}_{[010]}\tau_{1/2}$ and $\mathcal{M}_{[100]}\tau_{1/2}$ (here $\mathcal{M}$ and $\tau$ denote mirror and translation operations, respectively) from $P4_{2^{\prime}}/mnm^{\prime}$ magnetic space group.  The vanishing value of the $xy$ component of transport coefficients for $\theta = (\mathbb{N}+1/2)\pi$ is due to the $\mathcal{TM}_{[001]}$ symmetry from $Cmm^{\prime}m^{\prime}$ magnetic space group.
	
	Figures~\ref{fig2}(a)-\ref{fig2}(c) depict three prominent features of crystal thermal transport. Firstly, $zx$ and $yz$ components are nearly identical. This is ensured  by the symmetry constraint $\mathcal{TC}_{2x\bar{y}} R_{yz} = R_{zx}$ [see Eq.~\eqref{eq:R}], here, $\mathcal{C}_{2x\bar{y}}$ forces $R_{yz}$ to $-R_{zx}$, and $R_{zx}$ is odd under $\mathcal{T}$.  Secondly, all components show an oscillatory behavior with $\theta$: while the $zx$ and $yz$ components exhibit a period of 2$\pi$, $\sigma_{xy}$ is $\pi$-periodic. This can be understood  from realizing that the $z$-component of the spins at $\pi-\theta$ is the time-reversed counterpart of the ones at $\theta$ with $x$ and $y$ components staying the same, which results in the fact that $R_{zx,yz} (\pi-\theta)=R_{zx,yz} (\theta)$ but $R_{xy} (\pi-\theta)=-R_{xy} (\theta)$.  Finally, $\sigma$, $\alpha$, and $\kappa$ are strongly anisotropic in the direction of the N{\'e}el vector. For example, $zx$ and $yz$ components gradually increase with increasing $\theta$ from $0^{\circ}$ to $90^{\circ}$, which can be explained from the evolution of Berry curvature (supplemental Fig.~\textcolor{blue}{S1}).
	
\begin{figure}
		\centering
		\includegraphics[width=\columnwidth]{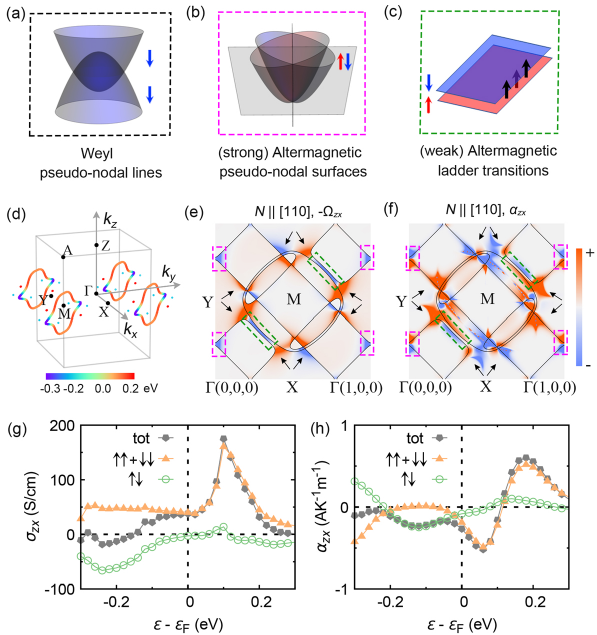}
		\caption{Three distinct types of interband transitions in altermagnetic RuO$_2$: (a) pseudo-nodal lines forming among the bands of the same spin, (b) altermagnetic pseudo-nodal surfaces forming among the bands of opposite spin driven by crystal symmetry, and (c) ladder transitions among weakly split bands of opposite spin with similar dispersion.  (d) Momentum and energy distribution of topological nodal lines and Weyl points.  (e) Relativistic Fermi surface (black lines) and Berry curvature $\Omega_{zx}$ (color maps, in atomic units) on the (001) plane at the true Fermi energy.  The contributions from gapped nodal lines, pseudo-nodal surfaces, and ladder transitions are indicated by black arrows, pink and green dashed rectangles, respectively.  (f) Similar to (e) but for the anomalous Nernst conductivity $\alpha_{zx}$.  (g,h) Total $\sigma_{zx}$ and $\alpha_{zx}$ and their decompositions to spin conserved ($\uparrow\uparrow+\downarrow\downarrow$) and spin flip ($\uparrow\downarrow$) parts.  In (d-h), the N{\'e}el vector points to \textbf{\textit{N}}($\varphi=45^\circ, \theta=90^\circ$).}
		\label{fig3}
	\end{figure}

	We now illustrate three distinct types of geometrical Berry curvature contributions that shape the crystal electronic and thermal transport properties in altermagnets: Weyl fermions, pseudo-nodal surfaces, and ladder transitions [Figs.~\ref{fig3}(a-c)]. As $\theta$ increases, the band splitting magnitude at specific $k$-points diminishes, and gaps along X-M-Y and Z-M-A paths progressively close [Fig.~\ref{fig2}(d)], signaling the emergence of N{\'e}el-order-dependent topological Weyl nodal features~\cite{Smejkal2020,Ahn2019,J-Zhan2023}. Without SOC, we identify two spin-up nodal lines on the (001) and (110) planes and two spin-down nodal lines on the (001) and ($\bar{1}10$) planes, connected by $\mathcal{C}_{4z}$ rotation symmetry (Fig.~\textcolor{blue}{S2}). Upon incorporating SOC, for $\textbf{\textit{N}}\parallel[110]$, the nodal line on the (110) plane is preserved, protected by $\mathcal{M}_{[110]}$ symmetry, while those on the (001) plane disperse into three pairs of Weyl points [Figs.~\ref{fig3}(d) and~\textcolor{blue}{S2}]. In Figs.~\ref{fig3}(e) and~\textcolor{blue}{S3}(a), we show how large Berry curvature emerges in regions of Weyl points due to interband transitions within bands of the same spin (spin-conserved parts, $\uparrow\uparrow$ or $\downarrow\downarrow$), schematically depicted in Fig.~\ref{fig3}(a).  The Weyl point contribution becomes more pronounced at the Fermi energy $\varepsilon = \varepsilon_F \pm 0.19$ eV [Figs.~\textcolor{blue}{S3}(b,c)].
	
	On the other hand, due to spin group symmetries $\mathcal{TS}_{2x}$ and $\mathcal{TS}_{2y}$ ($\mathcal{S}_{2x,2y}$ being twofold screw rotations), the degeneracy of spin-up and spin-down states on the $k_x=0, \pi$ and $k_y=0, \pi$ planes is enforced [e.g., $\Gamma$Y and $\Gamma$X paths in Fig.~\ref{fig1}(e)], creating four nodal planes [Fig.~\textcolor{blue}{S3}(d)], as also observed for FeSb$_2$~\cite{Mazin2021} and CaFeO$_3$\cite{RW-Zhang2021}. This degeneracy is lifted by SOC, resulting in substantial Berry curvature near these pseudo-nodal surfaces [Figs.~\ref{fig3}(e) and~\textcolor{blue}{S3}(a)]. The Berry curvature's origin here is tied to interband transitions between states of opposite spin, activated near high-symmetry nodal planes, Fig.~\ref{fig3}(b). Remarkably, the Berry curvature also arises from regions where a very small gap between spin-up and spin-down bands exists without SOC~\cite{note1}. This stems from spin-flip transitions between two spin-polarized bands exhibiting similar dispersion across the Fermi energy [Fig.~\ref{fig3}(c)], visible as rings centered at the M point in Fig.~\ref{fig1}(e). Such ``ladder" transitions have been observed in the anomalous Hall effect of ferromagnetic FePt alloys~\cite{HB-Zhang2011}. The distribution of the Berry curvature from ladder transitions in RuO$_2$ appears asymmetric along different $\Gamma$M paths when $\textbf{\textit{N}}\parallel[110]$ [Figs.~\ref{fig3}(e) and~\textcolor{blue}{S3}(a)], while symmetric patterns emerge when $\textbf{\textit{N}}\parallel[100]$ or $[010]$ (Fig.~\textcolor{blue}{S4}).
	
	We emphasize that among the three types of Berry curvature we considered, the latter two exhibit a distinctive feature of altermagnetism. These are generated by crystal symmetry-driven (quasi-)degeneracies between states with opposite spin, which foster the spin-flip ($\uparrow\downarrow$) transitions. Specifically, at the true Fermi energy, the $k$-resolved distribution of Nernst conductivity showcases significant contributions from Weyl points [Figs.~\ref{fig3}(f) and~\textcolor{blue}{S5}(c,d)]. However, with a slight adjustment to the position of the Fermi energy, the Nernst and Hall conductivities display substantial spin-flip contributions [Figs.~\ref{fig3}(g,h)].  Notably, in the regime of hole-doping (e.g., -0.2$\sim$0.0 eV), the ANE is entirely dominated by the spin-flip process. This starkly contrasts with conventional collinear antiferromagnets with $\mathcal{TS}$ symmetry, such as bilayer MnBi$_2$Te$_4$, in which the contributions from spin-conserved parts ($\uparrow\uparrow$ and $\downarrow\downarrow$) tend to cancel each other, and the impact of the spin-flip part is negligible (Fig.~\textcolor{blue}{S6}).
	
	Next, we turn our attention to the magnitude of crystal thermal transport coefficients. Figures~\ref{fig2}(e)-\ref{fig2}(g) show $\sigma$, $\alpha$, and $\kappa$ as a function of the Fermi energy for different temperatures.  At the true Fermi energy ($\varepsilon = \varepsilon_F$), $\sigma$ stays nearly constant with the increasing of temperature, while $\alpha$ and $\kappa$ change drastically upon heating.  In addition, although $\sigma$ is rather small,  $\alpha$ and $\kappa$ can reach as much as $-0.35$ AK$^{-1}$m$^{-1}$ and 5.5$\times 10^{-2}$ WK$^{-1}$m$^{-1}$~\cite{note2}. These values can be further substantially enhanced by engineering the degree of band filling via electron or hole doping.
	
	In understanding the temperature dependence of thermal transport let us first turn to the low-temperature limit.  Here, $\sigma$, $\alpha$, and $\kappa$ [Eq.~\eqref{eq:kappa}] are reduced to the first-order correction in the Sommerfeld expansion~\cite{Ashcroft1976}: \begin{eqnarray}
		\sigma_{ij} &\approx &\sigma_{ij}^{T = 0}(\mu), \label{sigma2}\\ 
		\alpha_{ij} &\approx &  -\frac{\pi^2 k_B^2T}{3e} \left.\frac{\textnormal{d} \sigma_{ij}^{T = 0}(\varepsilon)}{\textnormal{d} \varepsilon}\right|_{\varepsilon = \mu}, \label{alpha2}\\ 
		\kappa_{ij} &\approx  &  \frac{\pi^2 k_B^2 T}{3 e^2} \sigma_{ij}^{T = 0}(\mu), \label{kappa2}
	\end{eqnarray}
	where $e$ is the elementary charge and $k_B$ is the Boltzmann constant. Relations~\eqref{alpha2} and~\eqref{kappa2} reflect two well-known expressions: one is the Mott relation, linking the ANC to the energy derivative of AHC; another is the WF law which introduces the anomalous Lorenz ratio ${L}_{ij} = \kappa_{ij}/(\sigma_{ij}T)$, converging to the Sommerfeld constant (${L}_{0} =   \pi^2 k_{B}^{2} /(3e^2)= 2.44 \times 10^{-8}\ \Omega \textrm{WK}^{-2}$) in the low-temperature limit.  %Below, we assess the validity regions of low-temperature physics in RuO$_2$.
	
	\begin{figure}
		\centering
		\includegraphics[width=0.9\columnwidth]{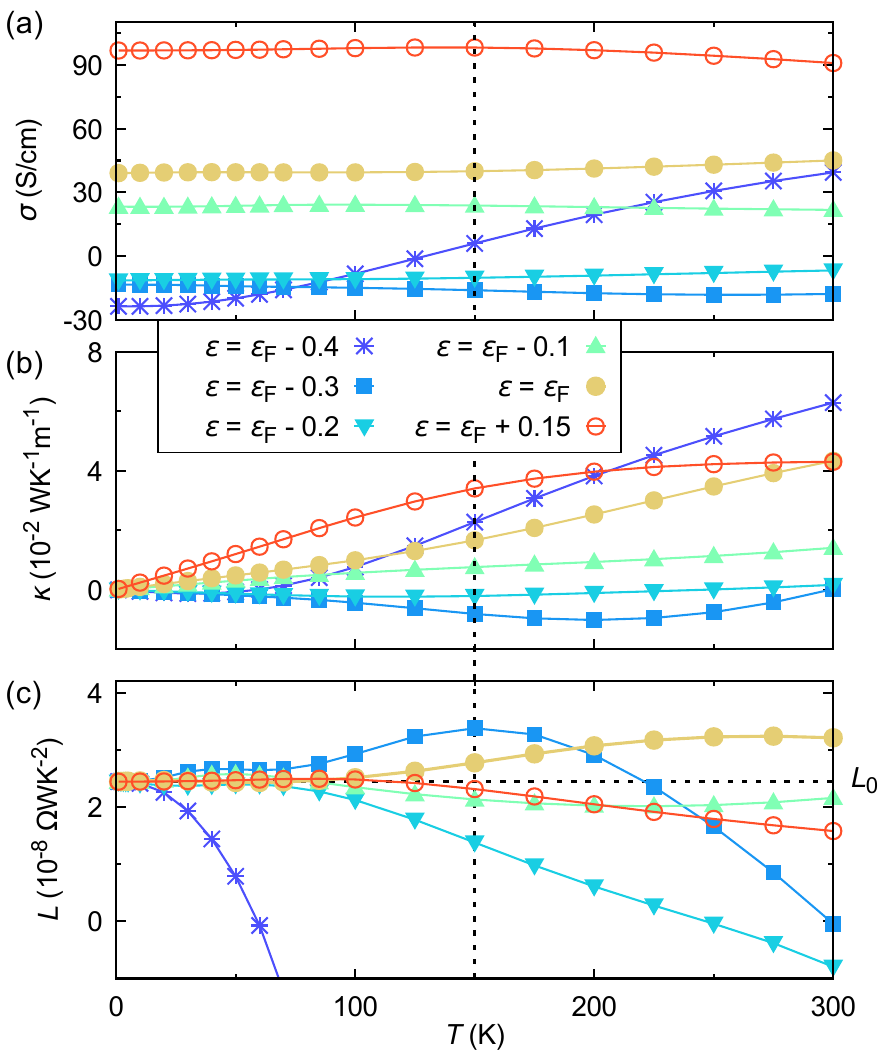}
		\caption{Temperature-dependence of (a) anomalous Hall conductivity $\sigma$, (b) anomalous thermal Hall conductivity $\kappa$, and (c) anomalous Lorenz ratio $L$ for $\textbf{\textit{N}}(\varphi=45^\circ, \theta=90^\circ)$ with different Fermi energies $\varepsilon$.  The horizontal dashed line in (c) denotes the Sommerfeld constant $L_{0}$.  The vertical dashed line denotes the allowed maximal temperature range of the Wiedemann-Franz law.}
		\label{fig4}
	\end{figure}
	
	The variation of $\sigma$, $\kappa$, and $L$ with temperature for different Fermi energies is shown in Fig.~\ref{fig4}.  When $\varepsilon = \varepsilon_F$, the anomalous Lorenz ratio $L$ is close to the Sommerfeld constant $L_0$ for $T< 100$ K [Fig.~\ref{fig4}(c)]. This is because in the low-temperature region, $L$ is dominated by $\sigma$ due to the linear dependence of $\kappa$ on temperature [Fig.~\ref{fig4}(b)], and $\sigma$ is nearly constant below 100 K [Fig.~\ref{fig4}(a)].  The WF law is valid in the energy range of the nodal line ($-0.2 \sim 0.2$ eV), and its robustness gradually weakens when going away from this energy range. Particularly, with the energy approaching the crossing points at about $\varepsilon = \varepsilon_F + 0.15$ eV [Fig.~\ref{fig2}(d)], the WF law is valid even up to 150 K. Hence, the robust validity of the WF law in RuO$_2$ at relatively high temperatures is likely related to its Weyl nodal properties, which have a topological origin.  The underlying physics presents a stark contrast to the expectations based on typical experiences with conventional ferromagnets~\cite{Onose2008,Shiomi2009,Shiomi2010}.
	
	Here, we do not consider the effect of inelastic scattering due to phonons~\cite{Strohm2005,Saito2019}, magnons~\cite{Onose2010}, and their interactions~\cite{X-Zhang2019}, which usually break the validity of anomalous WF law for finite temperatures, as observed in traditional ferromagnets~\cite{Onose2008,Shiomi2009,Shiomi2010}. Recently, the anomalous WF law is found to be valid in a wide temperature range for noncollinear AFMs, promoting the unique role of the intrinsic mechanism for thermal properties~\cite{XK-Li2017,LC-Xu2020,Sugii2019}. The intrinsic contribution can be further significantly enhanced by tuning the Fermi energy to lie at the position of magnetic topological (quasi-)degeneracies. It strongly indicates that the dominant intrinsic mechanism in RuO$_2$ comes from its novel topological nodal features.  We suggest that the unconventional anomalous thermal effects in RuO$_2$ can be observed experimentally in the [110] or [001] oriented films [Figs.~\ref{fig1}(a) and~\ref{fig1}(b)], as in the case of previously reported anomalous Hall measurements~\cite{ZX-Feng2022}.
	
	In this study, we have employed RuO$_2$ as a prototype to explore the crystal thermal transport of altermagnets, while our findings hold broader relevance across a diverse spectrum of altermagnets. For instance, the presence of a nonzero spin-flip process, a key feature of crystal thermal transport, is distinctly observable in another altermagnet, MnTe~\cite{Krempasky2023} (Fig.~\textcolor{blue}{S7}). Overall, the unexpected crystal thermal properties in room-temperature altermagnets, RuO$_2$ and MnTe, are poised to generate significant interest in the field of altermagnetic spin-caloritronics.

	\begin{acknowledgments}
This work is supported by the National Key R\&D
Program of China (Grant No. 2022YFA1402600 and
No. 2022YFA1403800), the Joint Sino-German Research
Projects (Chinese Grant No. 12061131002 and German
Research Foundation, DFG, Grant No. 44880005), the
Sino-German Mobility Programme (Grant No. M-0142),
the National Natural Science Foundation of China (Grant
No. 12274027, No. 12234003, No. 12321004, and
No. 12304066), the Natural Science Foundation of
Jiangsu Province (Grant No. BK20230684). Y. M., L. S.
and J. S. acknowledge support by the DFG-TRR 288-
422213477 and DFG-TRR 173/2-268565370.
	\end{acknowledgments}

%\bibliography{./ref.bib}

\begin{thebibliography}{0}%
\makeatletter
\providecommand \@ifxundefined [1]{%
 \@ifx{#1\undefined}
}%
\providecommand \@ifnum [1]{%
 \ifnum #1\expandafter \@firstoftwo
 \else \expandafter \@secondoftwo
 \fi
}%
\providecommand \@ifx [1]{%
 \ifx #1\expandafter \@firstoftwo
 \else \expandafter \@secondoftwo
 \fi
}%
\providecommand \natexlab [1]{#1}%
\providecommand \enquote  [1]{``#1''}%
\providecommand \bibnamefont  [1]{#1}%
\providecommand \bibfnamefont [1]{#1}%
\providecommand \citenamefont [1]{#1}%
\providecommand \href@noop [0]{\@secondoftwo}%
\providecommand \href [0]{\begingroup \@sanitize@url \@href}%
\providecommand \@href[1]{\@@startlink{#1}\@@href}%
\providecommand \@@href[1]{\endgroup#1\@@endlink}%
\providecommand \@sanitize@url [0]{\catcode `\\12\catcode `\$12\catcode
  `\&12\catcode `\#12\catcode `\^12\catcode `\_12\catcode `\%12\relax}%
\providecommand \@@startlink[1]{}%
\providecommand \@@endlink[0]{}%
\providecommand \url  [0]{\begingroup\@sanitize@url \@url }%
\providecommand \@url [1]{\endgroup\@href {#1}{\urlprefix }}%
\providecommand \urlprefix  [0]{URL }%
\providecommand \Eprint [0]{\href }%
\providecommand \doibase [0]{http://dx.doi.org/}%
\providecommand \selectlanguage [0]{\@gobble}%
\providecommand \bibinfo  [0]{\@secondoftwo}%
\providecommand \bibfield  [0]{\@secondoftwo}%
\providecommand \translation [1]{[#1]}%
\providecommand \BibitemOpen [0]{}%
\providecommand \bibitemStop [0]{}%
\providecommand \bibitemNoStop [0]{.\EOS\space}%
\providecommand \EOS [0]{\spacefactor3000\relax}%
\providecommand \BibitemShut  [1]{\csname bibitem#1\endcsname}%
\let\auto@bib@innerbib\@empty
%</preamble>
\end{thebibliography}%


\begin{thebibliography}{70}%
\makeatletter
\providecommand \@ifxundefined [1]{%
 \@ifx{#1\undefined}
}%
\providecommand \@ifnum [1]{%
 \ifnum #1\expandafter \@firstoftwo
 \else \expandafter \@secondoftwo
 \fi
}%
\providecommand \@ifx [1]{%
 \ifx #1\expandafter \@firstoftwo
 \else \expandafter \@secondoftwo
 \fi
}%
\providecommand \natexlab [1]{#1}%
\providecommand \enquote  [1]{``#1''}%
\providecommand \bibnamefont  [1]{#1}%
\providecommand \bibfnamefont [1]{#1}%
\providecommand \citenamefont [1]{#1}%
\providecommand \href@noop [0]{\@secondoftwo}%
\providecommand \href [0]{\begingroup \@sanitize@url \@href}%
\providecommand \@href[1]{\@@startlink{#1}\@@href}%
\providecommand \@@href[1]{\endgroup#1\@@endlink}%
\providecommand \@sanitize@url [0]{\catcode `\\12\catcode `\$12\catcode
  `\&12\catcode `\#12\catcode `\^12\catcode `\_12\catcode `\%12\relax}%
\providecommand \@@startlink[1]{}%
\providecommand \@@endlink[0]{}%
\providecommand \url  [0]{\begingroup\@sanitize@url \@url }%
\providecommand \@url [1]{\endgroup\@href {#1}{\urlprefix }}%
\providecommand \urlprefix  [0]{URL }%
\providecommand \Eprint [0]{\href }%
\providecommand \doibase [0]{http://dx.doi.org/}%
\providecommand \selectlanguage [0]{\@gobble}%
\providecommand \bibinfo  [0]{\@secondoftwo}%
\providecommand \bibfield  [0]{\@secondoftwo}%
\providecommand \translation [1]{[#1]}%
\providecommand \BibitemOpen [0]{}%
\providecommand \bibitemStop [0]{}%
\providecommand \bibitemNoStop [0]{.\EOS\space}%
\providecommand \EOS [0]{\spacefactor3000\relax}%
\providecommand \BibitemShut  [1]{\csname bibitem#1\endcsname}%
\let\auto@bib@innerbib\@empty
%</preamble>
\bibitem [{\citenamefont {Xiao}\ \emph {et~al.}(2006)\citenamefont {Xiao},
  \citenamefont {Yao}, \citenamefont {Fang},\ and\ \citenamefont
  {Niu}}]{D-Xiao2006}%
  \BibitemOpen
  \bibfield  {author} {\bibinfo {author} {\bibfnamefont {D.}~\bibnamefont
  {Xiao}}, \bibinfo {author} {\bibfnamefont {Y.}~\bibnamefont {Yao}}, \bibinfo
  {author} {\bibfnamefont {Z.}~\bibnamefont {Fang}}, \ and\ \bibinfo {author}
  {\bibfnamefont {Q.}~\bibnamefont {Niu}},\ }\href {\doibase
  10.1103/PhysRevLett.97.026603} {\bibfield  {journal} {\bibinfo  {journal}
  {Phys. Rev. Lett.}\ }\textbf {\bibinfo {volume} {97}},\ \bibinfo {pages}
  {026603} (\bibinfo {year} {2006})}\BibitemShut {NoStop}%
\bibitem [{\citenamefont {Qin}\ \emph {et~al.}(2011)\citenamefont {Qin},
  \citenamefont {Niu},\ and\ \citenamefont {Shi}}]{T-Qin2011}%
  \BibitemOpen
  \bibfield  {author} {\bibinfo {author} {\bibfnamefont {T.}~\bibnamefont
  {Qin}}, \bibinfo {author} {\bibfnamefont {Q.}~\bibnamefont {Niu}}, \ and\
  \bibinfo {author} {\bibfnamefont {J.}~\bibnamefont {Shi}},\ }\href {\doibase
  10.1103/PhysRevLett.107.236601} {\bibfield  {journal} {\bibinfo  {journal}
  {Phys. Rev. Lett.}\ }\textbf {\bibinfo {volume} {107}},\ \bibinfo {pages}
  {236601} (\bibinfo {year} {2011})}\BibitemShut {NoStop}%
\bibitem [{\citenamefont {Bauer}\ \emph {et~al.}(2012)\citenamefont {Bauer},
  \citenamefont {Saitoh},\ and\ \citenamefont {van Wees}}]{Bauer2012}%
  \BibitemOpen
  \bibfield  {author} {\bibinfo {author} {\bibfnamefont {G.~E.~W.}\
  \bibnamefont {Bauer}}, \bibinfo {author} {\bibfnamefont {E.}~\bibnamefont
  {Saitoh}}, \ and\ \bibinfo {author} {\bibfnamefont {B.~J.}\ \bibnamefont {van
  Wees}},\ }\href {\doibase 10.1038/nmat3301} {\bibfield  {journal} {\bibinfo
  {journal} {Nat. Mater.}\ }\textbf {\bibinfo {volume} {11}},\ \bibinfo {pages}
  {391} (\bibinfo {year} {2012})}\BibitemShut {NoStop}%
\bibitem [{\citenamefont {Boona}\ \emph {et~al.}(2014)\citenamefont {Boona},
  \citenamefont {Myers},\ and\ \citenamefont {Heremans}}]{Boona2014}%
  \BibitemOpen
  \bibfield  {author} {\bibinfo {author} {\bibfnamefont {S.~R.}\ \bibnamefont
  {Boona}}, \bibinfo {author} {\bibfnamefont {R.~C.}\ \bibnamefont {Myers}}, \
  and\ \bibinfo {author} {\bibfnamefont {J.~P.}\ \bibnamefont {Heremans}},\
  }\href {\doibase 10.1039/C3EE43299H} {\bibfield  {journal} {\bibinfo
  {journal} {Energy Environ. Sci.}\ }\textbf {\bibinfo {volume} {7}},\ \bibinfo
  {pages} {885} (\bibinfo {year} {2014})}\BibitemShut {NoStop}%
\bibitem [{\citenamefont {Callen}(1948)}]{Callen1948}%
  \BibitemOpen
  \bibfield  {author} {\bibinfo {author} {\bibfnamefont {H.~B.}\ \bibnamefont
  {Callen}},\ }\href {https://doi.org/10.1103/PhysRev.73.1349} {\bibfield
  {journal} {\bibinfo  {journal} {Phys. Rev.}\ }\textbf {\bibinfo {volume}
  {73}},\ \bibinfo {pages} {1349} (\bibinfo {year} {1948})}\BibitemShut
  {NoStop}%
\bibitem [{\citenamefont {Onose}\ \emph {et~al.}(2008)\citenamefont {Onose},
  \citenamefont {Shiomi},\ and\ \citenamefont {Tokura}}]{Onose2008}%
  \BibitemOpen
  \bibfield  {author} {\bibinfo {author} {\bibfnamefont {Y.}~\bibnamefont
  {Onose}}, \bibinfo {author} {\bibfnamefont {Y.}~\bibnamefont {Shiomi}}, \
  and\ \bibinfo {author} {\bibfnamefont {Y.}~\bibnamefont {Tokura}},\ }\href
  {\doibase 10.1103/PhysRevLett.100.016601} {\bibfield  {journal} {\bibinfo
  {journal} {Phys. Rev. Lett.}\ }\textbf {\bibinfo {volume} {100}},\ \bibinfo
  {pages} {016601} (\bibinfo {year} {2008})}\BibitemShut {NoStop}%
\bibitem [{\citenamefont {Shiomi}\ \emph {et~al.}(2009)\citenamefont {Shiomi},
  \citenamefont {Onose},\ and\ \citenamefont {Tokura}}]{Shiomi2009}%
  \BibitemOpen
  \bibfield  {author} {\bibinfo {author} {\bibfnamefont {Y.}~\bibnamefont
  {Shiomi}}, \bibinfo {author} {\bibfnamefont {Y.}~\bibnamefont {Onose}}, \
  and\ \bibinfo {author} {\bibfnamefont {Y.}~\bibnamefont {Tokura}},\ }\href
  {\doibase 10.1103/PhysRevB.79.100404} {\bibfield  {journal} {\bibinfo
  {journal} {Phys. Rev. B}\ }\textbf {\bibinfo {volume} {79}},\ \bibinfo
  {pages} {100404} (\bibinfo {year} {2009})}\BibitemShut {NoStop}%
\bibitem [{\citenamefont {Shiomi}\ \emph {et~al.}(2010)\citenamefont {Shiomi},
  \citenamefont {Onose},\ and\ \citenamefont {Tokura}}]{Shiomi2010}%
  \BibitemOpen
  \bibfield  {author} {\bibinfo {author} {\bibfnamefont {Y.}~\bibnamefont
  {Shiomi}}, \bibinfo {author} {\bibfnamefont {Y.}~\bibnamefont {Onose}}, \
  and\ \bibinfo {author} {\bibfnamefont {Y.}~\bibnamefont {Tokura}},\ }\href
  {https://doi.org/10.1103/PhysRevB.81.054414} {\bibfield  {journal} {\bibinfo
  {journal} {Phys. Rev. B}\ }\textbf {\bibinfo {volume} {81}},\ \bibinfo
  {pages} {054414} (\bibinfo {year} {2010})}\BibitemShut {NoStop}%
\bibitem [{\citenamefont {Smith}(1911)}]{Smith1911}%
  \BibitemOpen
  \bibfield  {author} {\bibinfo {author} {\bibfnamefont {A.~W.}\ \bibnamefont
  {Smith}},\ }\href {\doibase 10.1103/PhysRevSeriesI.33.295} {\bibfield
  {journal} {\bibinfo  {journal} {Phys. Rev. (Series I)}\ }\textbf {\bibinfo
  {volume} {33}},\ \bibinfo {pages} {295} (\bibinfo {year} {1911})}\BibitemShut
  {NoStop}%
\bibitem [{\citenamefont {Smith}(1921)}]{Smith1921}%
  \BibitemOpen
  \bibfield  {author} {\bibinfo {author} {\bibfnamefont {A.~W.}\ \bibnamefont
  {Smith}},\ }\href {\doibase 10.1103/PhysRev.17.23} {\bibfield  {journal}
  {\bibinfo  {journal} {Phys. Rev}\ }\textbf {\bibinfo {volume} {17}},\
  \bibinfo {pages} {23} (\bibinfo {year} {1921})}\BibitemShut {NoStop}%
\bibitem [{\citenamefont {Lee}\ \emph {et~al.}(2004)\citenamefont {Lee},
  \citenamefont {Watauchi}, \citenamefont {Miller}, \citenamefont {Cava},\ and\
  \citenamefont {Ong}}]{Lee2004}%
  \BibitemOpen
  \bibfield  {author} {\bibinfo {author} {\bibfnamefont {W.-L.}\ \bibnamefont
  {Lee}}, \bibinfo {author} {\bibfnamefont {S.}~\bibnamefont {Watauchi}},
  \bibinfo {author} {\bibfnamefont {V.~L.}\ \bibnamefont {Miller}}, \bibinfo
  {author} {\bibfnamefont {R.~J.}\ \bibnamefont {Cava}}, \ and\ \bibinfo
  {author} {\bibfnamefont {N.~P.}\ \bibnamefont {Ong}},\ }\href {\doibase
  10.1103/PhysRevLett.93.226601} {\bibfield  {journal} {\bibinfo  {journal}
  {Phys. Rev. Lett.}\ }\textbf {\bibinfo {volume} {93}},\ \bibinfo {pages}
  {226601} (\bibinfo {year} {2004})}\BibitemShut {NoStop}%
\bibitem [{\citenamefont {Miyasato}\ \emph {et~al.}(2007)\citenamefont
  {Miyasato}, \citenamefont {Abe}, \citenamefont {Fujii}, \citenamefont
  {Asamitsu}, \citenamefont {Onoda}, \citenamefont {Onose}, \citenamefont
  {Nagaosa},\ and\ \citenamefont {Tokura}}]{Miyasato2007}%
  \BibitemOpen
  \bibfield  {author} {\bibinfo {author} {\bibfnamefont {T.}~\bibnamefont
  {Miyasato}}, \bibinfo {author} {\bibfnamefont {N.}~\bibnamefont {Abe}},
  \bibinfo {author} {\bibfnamefont {T.}~\bibnamefont {Fujii}}, \bibinfo
  {author} {\bibfnamefont {A.}~\bibnamefont {Asamitsu}}, \bibinfo {author}
  {\bibfnamefont {S.}~\bibnamefont {Onoda}}, \bibinfo {author} {\bibfnamefont
  {Y.}~\bibnamefont {Onose}}, \bibinfo {author} {\bibfnamefont
  {N.}~\bibnamefont {Nagaosa}}, \ and\ \bibinfo {author} {\bibfnamefont
  {Y.}~\bibnamefont {Tokura}},\ }\href {\doibase 10.1103/PhysRevLett.99.086602}
  {\bibfield  {journal} {\bibinfo  {journal} {Phys. Rev. Lett.}\ }\textbf
  {\bibinfo {volume} {99}},\ \bibinfo {eid} {086602} (\bibinfo {year}
  {2007})}\BibitemShut {NoStop}%
\bibitem [{\citenamefont {Onoda}\ \emph {et~al.}(2008)\citenamefont {Onoda},
  \citenamefont {Sugimoto},\ and\ \citenamefont {Nagaosa}}]{Onoda2008}%
  \BibitemOpen
  \bibfield  {author} {\bibinfo {author} {\bibfnamefont {S.}~\bibnamefont
  {Onoda}}, \bibinfo {author} {\bibfnamefont {N.}~\bibnamefont {Sugimoto}}, \
  and\ \bibinfo {author} {\bibfnamefont {N.}~\bibnamefont {Nagaosa}},\ }\href
  {\doibase 10.1103/physrevb.77.165103} {\bibfield  {journal} {\bibinfo
  {journal} {Phys. Rev. B}\ }\textbf {\bibinfo {volume} {77}},\ \bibinfo
  {pages} {165103} (\bibinfo {year} {2008})}\BibitemShut {NoStop}%
\bibitem [{\citenamefont {Ikhlas}\ \emph {et~al.}(2017)\citenamefont {Ikhlas},
  \citenamefont {Tomita}, \citenamefont {Koretsune}, \citenamefont {Suzuki},
  \citenamefont {Nishio-Hamane}, \citenamefont {Arita}, \citenamefont {Otani},\
  and\ \citenamefont {Nakatsuji}}]{Ikhlas2017}%
  \BibitemOpen
  \bibfield  {author} {\bibinfo {author} {\bibfnamefont {M.}~\bibnamefont
  {Ikhlas}}, \bibinfo {author} {\bibfnamefont {T.}~\bibnamefont {Tomita}},
  \bibinfo {author} {\bibfnamefont {T.}~\bibnamefont {Koretsune}}, \bibinfo
  {author} {\bibfnamefont {M.-T.}\ \bibnamefont {Suzuki}}, \bibinfo {author}
  {\bibfnamefont {D.}~\bibnamefont {Nishio-Hamane}}, \bibinfo {author}
  {\bibfnamefont {R.}~\bibnamefont {Arita}}, \bibinfo {author} {\bibfnamefont
  {Y.}~\bibnamefont {Otani}}, \ and\ \bibinfo {author} {\bibfnamefont
  {S.}~\bibnamefont {Nakatsuji}},\ }\href {https://doi.org/10.1038/nphys4181}
  {\bibfield  {journal} {\bibinfo  {journal} {Nat. Phys.}\ }\textbf {\bibinfo
  {volume} {13}},\ \bibinfo {pages} {1085} (\bibinfo {year}
  {2017})}\BibitemShut {NoStop}%
\bibitem [{\citenamefont {Li}\ \emph {et~al.}(2017)\citenamefont {Li},
  \citenamefont {Xu}, \citenamefont {Ding}, \citenamefont {Wang}, \citenamefont
  {Shen}, \citenamefont {Lu}, \citenamefont {Zhu},\ and\ \citenamefont
  {Behnia}}]{XK-Li2017}%
  \BibitemOpen
  \bibfield  {author} {\bibinfo {author} {\bibfnamefont {X.}~\bibnamefont
  {Li}}, \bibinfo {author} {\bibfnamefont {L.}~\bibnamefont {Xu}}, \bibinfo
  {author} {\bibfnamefont {L.}~\bibnamefont {Ding}}, \bibinfo {author}
  {\bibfnamefont {J.}~\bibnamefont {Wang}}, \bibinfo {author} {\bibfnamefont
  {M.}~\bibnamefont {Shen}}, \bibinfo {author} {\bibfnamefont {X.}~\bibnamefont
  {Lu}}, \bibinfo {author} {\bibfnamefont {Z.}~\bibnamefont {Zhu}}, \ and\
  \bibinfo {author} {\bibfnamefont {K.}~\bibnamefont {Behnia}},\ }\href
  {\doibase 10.1103/PhysRevLett.119.056601} {\bibfield  {journal} {\bibinfo
  {journal} {Phys. Rev. Lett.}\ }\textbf {\bibinfo {volume} {119}},\ \bibinfo
  {pages} {056601} (\bibinfo {year} {2017})}\BibitemShut {NoStop}%
\bibitem [{\citenamefont {Guo}\ and\ \citenamefont {Wang}(2017)}]{GY-Guo2017}%
  \BibitemOpen
  \bibfield  {author} {\bibinfo {author} {\bibfnamefont {G.-Y.}\ \bibnamefont
  {Guo}}\ and\ \bibinfo {author} {\bibfnamefont {T.-C.}\ \bibnamefont {Wang}},\
  }\href {https://doi.org/10.1103/PhysRevB.96.224415} {\bibfield  {journal}
  {\bibinfo  {journal} {Phys. Rev. B}\ }\textbf {\bibinfo {volume} {96}},\
  \bibinfo {eid} {224415} (\bibinfo {year} {2017})}\BibitemShut {NoStop}%
\bibitem [{\citenamefont {Xu}\ \emph {et~al.}(2020)\citenamefont {Xu},
  \citenamefont {Li}, \citenamefont {Lu}, \citenamefont {Collignon},
  \citenamefont {Fu}, \citenamefont {Koo}, \citenamefont {Fauqu{\'{e}}},
  \citenamefont {Yan}, \citenamefont {Zhu},\ and\ \citenamefont
  {Behnia}}]{LC-Xu2020}%
  \BibitemOpen
  \bibfield  {author} {\bibinfo {author} {\bibfnamefont {L.}~\bibnamefont
  {Xu}}, \bibinfo {author} {\bibfnamefont {X.}~\bibnamefont {Li}}, \bibinfo
  {author} {\bibfnamefont {X.}~\bibnamefont {Lu}}, \bibinfo {author}
  {\bibfnamefont {C.}~\bibnamefont {Collignon}}, \bibinfo {author}
  {\bibfnamefont {H.}~\bibnamefont {Fu}}, \bibinfo {author} {\bibfnamefont
  {J.}~\bibnamefont {Koo}}, \bibinfo {author} {\bibfnamefont {B.}~\bibnamefont
  {Fauqu{\'{e}}}}, \bibinfo {author} {\bibfnamefont {B.}~\bibnamefont {Yan}},
  \bibinfo {author} {\bibfnamefont {Z.}~\bibnamefont {Zhu}}, \ and\ \bibinfo
  {author} {\bibfnamefont {K.}~\bibnamefont {Behnia}},\ }\href {\doibase
  10.1126/sciadv.aaz3522} {\bibfield  {journal} {\bibinfo  {journal} {Sci.
  Adv.}\ }\textbf {\bibinfo {volume} {6}},\ \bibinfo {pages} {eaaz3522}
  (\bibinfo {year} {2020})}\BibitemShut {NoStop}%
\bibitem [{\citenamefont {Sugii}\ \emph {et~al.}(2019)\citenamefont {Sugii},
  \citenamefont {Imai}, \citenamefont {Shimozawa}, \citenamefont {Ikhlas},
  \citenamefont {Kiyohara}, \citenamefont {Tomita}, \citenamefont {Suzuki},
  \citenamefont {Koretsune}, \citenamefont {Arita},\ and\ \citenamefont
  {Yamashita}}]{Sugii2019}%
  \BibitemOpen
  \bibfield  {author} {\bibinfo {author} {\bibfnamefont {K.}~\bibnamefont
  {Sugii}}, \bibinfo {author} {\bibfnamefont {Y.}~\bibnamefont {Imai}},
  \bibinfo {author} {\bibfnamefont {M.}~\bibnamefont {Shimozawa}}, \bibinfo
  {author} {\bibfnamefont {M.}~\bibnamefont {Ikhlas}}, \bibinfo {author}
  {\bibfnamefont {N.}~\bibnamefont {Kiyohara}}, \bibinfo {author}
  {\bibfnamefont {T.}~\bibnamefont {Tomita}}, \bibinfo {author} {\bibfnamefont
  {M.-T.}\ \bibnamefont {Suzuki}}, \bibinfo {author} {\bibfnamefont
  {T.}~\bibnamefont {Koretsune}}, \bibinfo {author} {\bibfnamefont {N.~S.}\
  \bibnamefont {Arita}, \bibfnamefont {Ryotaro}}, \ and\ \bibinfo {author}
  {\bibfnamefont {M.}~\bibnamefont {Yamashita}},\ }\href
  {https://arxiv.org/abs/1902.06601} {\bibfield  {journal} {\bibinfo  {journal}
  {arXiv: 1902.06601}\ } (\bibinfo {year} {2019})}\BibitemShut {NoStop}%
\bibitem [{\citenamefont {Zhou}\ \emph {et~al.}(2020)\citenamefont {Zhou},
  \citenamefont {Hanke}, \citenamefont {Feng}, \citenamefont {Bl{\"u}gel},
  \citenamefont {Mokrousov},\ and\ \citenamefont {Yao}}]{XD-Zhou2020}%
  \BibitemOpen
  \bibfield  {author} {\bibinfo {author} {\bibfnamefont {X.}~\bibnamefont
  {Zhou}}, \bibinfo {author} {\bibfnamefont {J.-P.}\ \bibnamefont {Hanke}},
  \bibinfo {author} {\bibfnamefont {W.}~\bibnamefont {Feng}}, \bibinfo {author}
  {\bibfnamefont {S.}~\bibnamefont {Bl{\"u}gel}}, \bibinfo {author}
  {\bibfnamefont {Y.}~\bibnamefont {Mokrousov}}, \ and\ \bibinfo {author}
  {\bibfnamefont {Y.}~\bibnamefont {Yao}},\ }\href
  {https://doi.org/10.1103/PhysRevMaterials.4.024408} {\bibfield  {journal}
  {\bibinfo  {journal} {Phys. Rev. Mater.}\ }\textbf {\bibinfo {volume} {4}},\
  \bibinfo {pages} {024408} (\bibinfo {year} {2020})}\BibitemShut {NoStop}%
\bibitem [{\citenamefont {Zhou}\ \emph {et~al.}(2019)\citenamefont {Zhou},
  \citenamefont {Hanke}, \citenamefont {Feng}, \citenamefont {Li},
  \citenamefont {Guo}, \citenamefont {Yao}, \citenamefont {Bl\"ugel},\ and\
  \citenamefont {Mokrousov}}]{XD-Zhou2019a}%
  \BibitemOpen
  \bibfield  {author} {\bibinfo {author} {\bibfnamefont {X.}~\bibnamefont
  {Zhou}}, \bibinfo {author} {\bibfnamefont {J.-P.}\ \bibnamefont {Hanke}},
  \bibinfo {author} {\bibfnamefont {W.}~\bibnamefont {Feng}}, \bibinfo {author}
  {\bibfnamefont {F.}~\bibnamefont {Li}}, \bibinfo {author} {\bibfnamefont
  {G.-Y.}\ \bibnamefont {Guo}}, \bibinfo {author} {\bibfnamefont
  {Y.}~\bibnamefont {Yao}}, \bibinfo {author} {\bibfnamefont {S.}~\bibnamefont
  {Bl\"ugel}}, \ and\ \bibinfo {author} {\bibfnamefont {Y.}~\bibnamefont
  {Mokrousov}},\ }\href {\doibase 10.1103/PhysRevB.99.104428} {\bibfield
  {journal} {\bibinfo  {journal} {Phys. Rev. B}\ }\textbf {\bibinfo {volume}
  {99}},\ \bibinfo {pages} {104428} (\bibinfo {year} {2019})}\BibitemShut
  {NoStop}%
\bibitem [{\citenamefont {Shiomi}\ \emph {et~al.}(2013)\citenamefont {Shiomi},
  \citenamefont {Kanazawa}, \citenamefont {Shibata}, \citenamefont {Onose},\
  and\ \citenamefont {Tokura}}]{Shiomi2013}%
  \BibitemOpen
  \bibfield  {author} {\bibinfo {author} {\bibfnamefont {Y.}~\bibnamefont
  {Shiomi}}, \bibinfo {author} {\bibfnamefont {N.}~\bibnamefont {Kanazawa}},
  \bibinfo {author} {\bibfnamefont {K.}~\bibnamefont {Shibata}}, \bibinfo
  {author} {\bibfnamefont {Y.}~\bibnamefont {Onose}}, \ and\ \bibinfo {author}
  {\bibfnamefont {Y.}~\bibnamefont {Tokura}},\ }\href {\doibase
  10.1103/physrevb.88.064409} {\bibfield  {journal} {\bibinfo  {journal} {Phys.
  Rev. B}\ }\textbf {\bibinfo {volume} {88}},\ \bibinfo {pages} {064409}
  (\bibinfo {year} {2013})}\BibitemShut {NoStop}%
\bibitem [{\citenamefont {Hirschberger}\ \emph {et~al.}(2020)\citenamefont
  {Hirschberger}, \citenamefont {Spitz}, \citenamefont {Nomoto}, \citenamefont
  {Kurumaji}, \citenamefont {Gao}, \citenamefont {Masell}, \citenamefont
  {Nakajima}, \citenamefont {Kikkawa}, \citenamefont {Yamasaki}, \citenamefont
  {Sagayama}, \citenamefont {Nakao}, \citenamefont {Taguchi}, \citenamefont
  {Arita}, \citenamefont {hisa Arima},\ and\ \citenamefont
  {Tokura}}]{Hirschberger2020}%
  \BibitemOpen
  \bibfield  {author} {\bibinfo {author} {\bibfnamefont {M.}~\bibnamefont
  {Hirschberger}}, \bibinfo {author} {\bibfnamefont {L.}~\bibnamefont {Spitz}},
  \bibinfo {author} {\bibfnamefont {T.}~\bibnamefont {Nomoto}}, \bibinfo
  {author} {\bibfnamefont {T.}~\bibnamefont {Kurumaji}}, \bibinfo {author}
  {\bibfnamefont {S.}~\bibnamefont {Gao}}, \bibinfo {author} {\bibfnamefont
  {J.}~\bibnamefont {Masell}}, \bibinfo {author} {\bibfnamefont
  {T.}~\bibnamefont {Nakajima}}, \bibinfo {author} {\bibfnamefont
  {A.}~\bibnamefont {Kikkawa}}, \bibinfo {author} {\bibfnamefont
  {Y.}~\bibnamefont {Yamasaki}}, \bibinfo {author} {\bibfnamefont
  {H.}~\bibnamefont {Sagayama}}, \bibinfo {author} {\bibfnamefont
  {H.}~\bibnamefont {Nakao}}, \bibinfo {author} {\bibfnamefont
  {Y.}~\bibnamefont {Taguchi}}, \bibinfo {author} {\bibfnamefont
  {R.}~\bibnamefont {Arita}}, \bibinfo {author} {\bibfnamefont
  {T.}~\bibnamefont {hisa Arima}}, \ and\ \bibinfo {author} {\bibfnamefont
  {Y.}~\bibnamefont {Tokura}},\ }\href {\doibase
  10.1103/physrevlett.125.076602} {\bibfield  {journal} {\bibinfo  {journal}
  {Phys. Rev. Lett.}\ }\textbf {\bibinfo {volume} {125}},\ \bibinfo {pages}
  {076602} (\bibinfo {year} {2020})}\BibitemShut {NoStop}%
\bibitem [{\citenamefont {Zhang}\ \emph
  {et~al.}(2021{\natexlab{a}})\citenamefont {Zhang}, \citenamefont {Xu},\ and\
  \citenamefont {Ke}}]{H-Zhang2021}%
  \BibitemOpen
  \bibfield  {author} {\bibinfo {author} {\bibfnamefont {H.}~\bibnamefont
  {Zhang}}, \bibinfo {author} {\bibfnamefont {C.~Q.}\ \bibnamefont {Xu}}, \
  and\ \bibinfo {author} {\bibfnamefont {X.}~\bibnamefont {Ke}},\ }\href
  {\doibase 10.1103/physrevb.103.l201101} {\bibfield  {journal} {\bibinfo
  {journal} {Phys. Rev. B}\ }\textbf {\bibinfo {volume} {103}},\ \bibinfo
  {pages} {l201101} (\bibinfo {year} {2021}{\natexlab{a}})}\BibitemShut
  {NoStop}%
\bibitem [{\citenamefont {Owerre}(2017)}]{Owerre2017}%
  \BibitemOpen
  \bibfield  {author} {\bibinfo {author} {\bibfnamefont {S.~A.}\ \bibnamefont
  {Owerre}},\ }\href {\doibase 10.1103/physrevb.95.014422} {\bibfield
  {journal} {\bibinfo  {journal} {Phys. Rev. B}\ }\textbf {\bibinfo {volume}
  {95}},\ \bibinfo {pages} {014422} (\bibinfo {year} {2017})}\BibitemShut
  {NoStop}%
\bibitem [{\citenamefont {Lu}\ \emph {et~al.}(2019)\citenamefont {Lu},
  \citenamefont {Guo}, \citenamefont {Koval},\ and\ \citenamefont
  {Jia}}]{YL-Lu2019}%
  \BibitemOpen
  \bibfield  {author} {\bibinfo {author} {\bibfnamefont {Y.}~\bibnamefont
  {Lu}}, \bibinfo {author} {\bibfnamefont {X.}~\bibnamefont {Guo}}, \bibinfo
  {author} {\bibfnamefont {V.}~\bibnamefont {Koval}}, \ and\ \bibinfo {author}
  {\bibfnamefont {C.}~\bibnamefont {Jia}},\ }\href {\doibase
  10.1103/physrevb.99.054409} {\bibfield  {journal} {\bibinfo  {journal} {Phys.
  Rev. B}\ }\textbf {\bibinfo {volume} {99}},\ \bibinfo {pages} {054409}
  (\bibinfo {year} {2019})}\BibitemShut {NoStop}%
\bibitem [{\citenamefont {Zhou}\ \emph {et~al.}(2016)\citenamefont {Zhou},
  \citenamefont {Liang}, \citenamefont {Weng}, \citenamefont {Chen},
  \citenamefont {Yao}, \citenamefont {Chen}, \citenamefont {Dong},\ and\
  \citenamefont {Guo}}]{J-Zhou2016}%
  \BibitemOpen
  \bibfield  {author} {\bibinfo {author} {\bibfnamefont {J.}~\bibnamefont
  {Zhou}}, \bibinfo {author} {\bibfnamefont {Q.-F.}\ \bibnamefont {Liang}},
  \bibinfo {author} {\bibfnamefont {H.}~\bibnamefont {Weng}}, \bibinfo {author}
  {\bibfnamefont {Y.~B.}\ \bibnamefont {Chen}}, \bibinfo {author}
  {\bibfnamefont {S.-H.}\ \bibnamefont {Yao}}, \bibinfo {author} {\bibfnamefont
  {Y.-F.}\ \bibnamefont {Chen}}, \bibinfo {author} {\bibfnamefont
  {J.}~\bibnamefont {Dong}}, \ and\ \bibinfo {author} {\bibfnamefont {G.-Y.}\
  \bibnamefont {Guo}},\ }\href
  {https://link.aps.org/doi/10.1103/PhysRevLett.116.256601} {\bibfield
  {journal} {\bibinfo  {journal} {Phys. Rev. Lett.}\ }\textbf {\bibinfo
  {volume} {116}},\ \bibinfo {pages} {256601} (\bibinfo {year}
  {2016})}\BibitemShut {NoStop}%
\bibitem [{\citenamefont {Hanke}\ \emph {et~al.}(2017)\citenamefont {Hanke},
  \citenamefont {Freimuth}, \citenamefont {Bl{\"u}gel},\ and\ \citenamefont
  {Mokrousov}}]{Hanke2017}%
  \BibitemOpen
  \bibfield  {author} {\bibinfo {author} {\bibfnamefont {J.-P.}\ \bibnamefont
  {Hanke}}, \bibinfo {author} {\bibfnamefont {F.}~\bibnamefont {Freimuth}},
  \bibinfo {author} {\bibfnamefont {S.}~\bibnamefont {Bl{\"u}gel}}, \ and\
  \bibinfo {author} {\bibfnamefont {Y.}~\bibnamefont {Mokrousov}},\ }\href
  {http://dx.doi.org/10.1038/srep41078} {\bibfield  {journal} {\bibinfo
  {journal} {Sci. Rep.}\ }\textbf {\bibinfo {volume} {7}},\ \bibinfo {pages}
  {41078} (\bibinfo {year} {2017})}\BibitemShut {NoStop}%
\bibitem [{\citenamefont {Feng}\ \emph {et~al.}(2020)\citenamefont {Feng},
  \citenamefont {Hanke}, \citenamefont {Zhou}, \citenamefont {Guo},
  \citenamefont {Bl{\"u}gel}, \citenamefont {Mokrousov},\ and\ \citenamefont
  {Yao}}]{WX-Feng2020}%
  \BibitemOpen
  \bibfield  {author} {\bibinfo {author} {\bibfnamefont {W.}~\bibnamefont
  {Feng}}, \bibinfo {author} {\bibfnamefont {J.-P.}\ \bibnamefont {Hanke}},
  \bibinfo {author} {\bibfnamefont {X.}~\bibnamefont {Zhou}}, \bibinfo {author}
  {\bibfnamefont {G.-Y.}\ \bibnamefont {Guo}}, \bibinfo {author} {\bibfnamefont
  {S.}~\bibnamefont {Bl{\"u}gel}}, \bibinfo {author} {\bibfnamefont
  {Y.}~\bibnamefont {Mokrousov}}, \ and\ \bibinfo {author} {\bibfnamefont
  {Y.}~\bibnamefont {Yao}},\ }\href
  {https://doi.org/10.1038/s41467-019-13968-8} {\bibfield  {journal} {\bibinfo
  {journal} {Nat. Commun.}\ }\textbf {\bibinfo {volume} {11}},\ \bibinfo
  {pages} {118} (\bibinfo {year} {2020})}\BibitemShut {NoStop}%
\bibitem [{\citenamefont {{\v S}mejkal}\ \emph {et~al.}(2020)\citenamefont {{\v
  S}mejkal}, \citenamefont {Gonz{\'a}lez-Hern{\'a}ndez}, \citenamefont
  {Jungwirth},\ and\ \citenamefont {Sinova}}]{Smejkal2020}%
  \BibitemOpen
  \bibfield  {author} {\bibinfo {author} {\bibfnamefont {L.}~\bibnamefont {{\v
  S}mejkal}}, \bibinfo {author} {\bibfnamefont {R.}~\bibnamefont
  {Gonz{\'a}lez-Hern{\'a}ndez}}, \bibinfo {author} {\bibfnamefont
  {T.}~\bibnamefont {Jungwirth}}, \ and\ \bibinfo {author} {\bibfnamefont
  {J.}~\bibnamefont {Sinova}},\ }\href {\doibase 10.1126/sciadv.aaz8809}
  {\bibfield  {journal} {\bibinfo  {journal} {Sci. Adv.}\ }\textbf {\bibinfo
  {volume} {6}},\ \bibinfo {pages} {eaaz8809} (\bibinfo {year}
  {2020})}\BibitemShut {NoStop}%
\bibitem [{\citenamefont {Feng}\ \emph {et~al.}(2022)\citenamefont {Feng},
  \citenamefont {Zhou}, \citenamefont {{\v S}mejkal}, \citenamefont {Wu},
  \citenamefont {Zhu}, \citenamefont {Guo}, \citenamefont
  {Gonz{\'a}lez-Hern{\'a}ndez}, \citenamefont {Wang}, \citenamefont {Yan},
  \citenamefont {Qin}, \citenamefont {Zhang}, \citenamefont {Wu}, \citenamefont
  {Chen}, \citenamefont {Meng}, \citenamefont {Liu}, \citenamefont {Xia},
  \citenamefont {Sinova}, \citenamefont {Jungwirth},\ and\ \citenamefont
  {Liu}}]{ZX-Feng2022}%
  \BibitemOpen
  \bibfield  {author} {\bibinfo {author} {\bibfnamefont {Z.}~\bibnamefont
  {Feng}}, \bibinfo {author} {\bibfnamefont {X.}~\bibnamefont {Zhou}}, \bibinfo
  {author} {\bibfnamefont {L.}~\bibnamefont {{\v S}mejkal}}, \bibinfo {author}
  {\bibfnamefont {L.}~\bibnamefont {Wu}}, \bibinfo {author} {\bibfnamefont
  {Z.}~\bibnamefont {Zhu}}, \bibinfo {author} {\bibfnamefont {H.}~\bibnamefont
  {Guo}}, \bibinfo {author} {\bibfnamefont {R.}~\bibnamefont
  {Gonz{\'a}lez-Hern{\'a}ndez}}, \bibinfo {author} {\bibfnamefont
  {X.}~\bibnamefont {Wang}}, \bibinfo {author} {\bibfnamefont {H.}~\bibnamefont
  {Yan}}, \bibinfo {author} {\bibfnamefont {P.}~\bibnamefont {Qin}}, \bibinfo
  {author} {\bibfnamefont {X.}~\bibnamefont {Zhang}}, \bibinfo {author}
  {\bibfnamefont {H.}~\bibnamefont {Wu}}, \bibinfo {author} {\bibfnamefont
  {H.}~\bibnamefont {Chen}}, \bibinfo {author} {\bibfnamefont {Z.}~\bibnamefont
  {Meng}}, \bibinfo {author} {\bibfnamefont {L.}~\bibnamefont {Liu}}, \bibinfo
  {author} {\bibfnamefont {Z.}~\bibnamefont {Xia}}, \bibinfo {author}
  {\bibfnamefont {J.}~\bibnamefont {Sinova}}, \bibinfo {author} {\bibfnamefont
  {T.}~\bibnamefont {Jungwirth}}, \ and\ \bibinfo {author} {\bibfnamefont
  {Z.}~\bibnamefont {Liu}},\ }\href {\doibase 10.1038/s41928-022-00866-z}
  {\bibfield  {journal} {\bibinfo  {journal} {Nat. Electron.}\ }\textbf
  {\bibinfo {volume} {5}},\ \bibinfo {pages} {735} (\bibinfo {year}
  {2022})}\BibitemShut {NoStop}%
\bibitem [{\citenamefont {\ifmmode~\check{S}\else \v{S}\fi{}mejkal}\ \emph
  {et~al.}(2022{\natexlab{a}})\citenamefont {\ifmmode~\check{S}\else
  \v{S}\fi{}mejkal}, \citenamefont {Sinova},\ and\ \citenamefont
  {Jungwirth}}]{Smejkal2022b}%
  \BibitemOpen
  \bibfield  {author} {\bibinfo {author} {\bibfnamefont {L.}~\bibnamefont
  {\ifmmode~\check{S}\else \v{S}\fi{}mejkal}}, \bibinfo {author} {\bibfnamefont
  {J.}~\bibnamefont {Sinova}}, \ and\ \bibinfo {author} {\bibfnamefont
  {T.}~\bibnamefont {Jungwirth}},\ }\href {\doibase 10.1103/PhysRevX.12.031042}
  {\bibfield  {journal} {\bibinfo  {journal} {Phys. Rev. X}\ }\textbf {\bibinfo
  {volume} {12}},\ \bibinfo {pages} {031042} (\bibinfo {year}
  {2022}{\natexlab{a}})}\BibitemShut {NoStop}%
\bibitem [{\citenamefont {\ifmmode~\check{S}\else \v{S}\fi{}mejkal}\ \emph
  {et~al.}(2022{\natexlab{b}})\citenamefont {\ifmmode~\check{S}\else
  \v{S}\fi{}mejkal}, \citenamefont {Sinova},\ and\ \citenamefont
  {Jungwirth}}]{Smejkal2022a}%
  \BibitemOpen
  \bibfield  {author} {\bibinfo {author} {\bibfnamefont {L.}~\bibnamefont
  {\ifmmode~\check{S}\else \v{S}\fi{}mejkal}}, \bibinfo {author} {\bibfnamefont
  {J.}~\bibnamefont {Sinova}}, \ and\ \bibinfo {author} {\bibfnamefont
  {T.}~\bibnamefont {Jungwirth}},\ }\href {\doibase 10.1103/PhysRevX.12.040501}
  {\bibfield  {journal} {\bibinfo  {journal} {Phys. Rev. X}\ }\textbf {\bibinfo
  {volume} {12}},\ \bibinfo {pages} {040501} (\bibinfo {year}
  {2022}{\natexlab{b}})}\BibitemShut {NoStop}%
\bibitem [{\citenamefont {Ma}\ \emph {et~al.}(2021)\citenamefont {Ma},
  \citenamefont {Hu}, \citenamefont {Li}, \citenamefont {Liu}, \citenamefont
  {Yao}, \citenamefont {Jia},\ and\ \citenamefont {Liu}}]{HY-Ma2021}%
  \BibitemOpen
  \bibfield  {author} {\bibinfo {author} {\bibfnamefont {H.-Y.}\ \bibnamefont
  {Ma}}, \bibinfo {author} {\bibfnamefont {M.}~\bibnamefont {Hu}}, \bibinfo
  {author} {\bibfnamefont {N.}~\bibnamefont {Li}}, \bibinfo {author}
  {\bibfnamefont {J.}~\bibnamefont {Liu}}, \bibinfo {author} {\bibfnamefont
  {W.}~\bibnamefont {Yao}}, \bibinfo {author} {\bibfnamefont {J.-F.}\
  \bibnamefont {Jia}}, \ and\ \bibinfo {author} {\bibfnamefont
  {J.}~\bibnamefont {Liu}},\ }\href {\doibase 10.1038/s41467-021-23127-7}
  {\bibfield  {journal} {\bibinfo  {journal} {Nat. Commun.}\ }\textbf {\bibinfo
  {volume} {12}},\ \bibinfo {pages} {2846} (\bibinfo {year}
  {2021})}\BibitemShut {NoStop}%
\bibitem [{\citenamefont {Berlijn}\ \emph {et~al.}(2017)\citenamefont
  {Berlijn}, \citenamefont {Snijders}, \citenamefont {Delaire}, \citenamefont
  {Zhou}, \citenamefont {Maier}, \citenamefont {Cao}, \citenamefont {Chi},
  \citenamefont {Matsuda}, \citenamefont {Wang}, \citenamefont {Koehler},
  \citenamefont {Kent},\ and\ \citenamefont {Weitering}}]{Berlijn2017}%
  \BibitemOpen
  \bibfield  {author} {\bibinfo {author} {\bibfnamefont {T.}~\bibnamefont
  {Berlijn}}, \bibinfo {author} {\bibfnamefont {P.~C.}\ \bibnamefont
  {Snijders}}, \bibinfo {author} {\bibfnamefont {O.}~\bibnamefont {Delaire}},
  \bibinfo {author} {\bibfnamefont {H.-D.}\ \bibnamefont {Zhou}}, \bibinfo
  {author} {\bibfnamefont {T.~A.}\ \bibnamefont {Maier}}, \bibinfo {author}
  {\bibfnamefont {H.-B.}\ \bibnamefont {Cao}}, \bibinfo {author} {\bibfnamefont
  {S.-X.}\ \bibnamefont {Chi}}, \bibinfo {author} {\bibfnamefont
  {M.}~\bibnamefont {Matsuda}}, \bibinfo {author} {\bibfnamefont
  {Y.}~\bibnamefont {Wang}}, \bibinfo {author} {\bibfnamefont {M.~R.}\
  \bibnamefont {Koehler}}, \bibinfo {author} {\bibfnamefont {P.~R.~C.}\
  \bibnamefont {Kent}}, \ and\ \bibinfo {author} {\bibfnamefont {H.~H.}\
  \bibnamefont {Weitering}},\ }\href {\doibase 10.1103/PhysRevLett.118.077201}
  {\bibfield  {journal} {\bibinfo  {journal} {Phys. Rev. Lett.}\ }\textbf
  {\bibinfo {volume} {118}},\ \bibinfo {eid} {077201} (\bibinfo {year}
  {2017})}\BibitemShut {NoStop}%
\bibitem [{\citenamefont {Zhu}\ \emph {et~al.}(2019)\citenamefont {Zhu},
  \citenamefont {Strempfer}, \citenamefont {Rao}, \citenamefont {Occhialini},
  \citenamefont {Pelliciari}, \citenamefont {Choi}, \citenamefont {Kawaguchi},
  \citenamefont {You}, \citenamefont {Mitchell}, \citenamefont {Shao-Horn},\
  and\ \citenamefont {Comin}}]{ZH-Zhu2019}%
  \BibitemOpen
  \bibfield  {author} {\bibinfo {author} {\bibfnamefont {Z.~H.}\ \bibnamefont
  {Zhu}}, \bibinfo {author} {\bibfnamefont {J.}~\bibnamefont {Strempfer}},
  \bibinfo {author} {\bibfnamefont {R.~R.}\ \bibnamefont {Rao}}, \bibinfo
  {author} {\bibfnamefont {C.~A.}\ \bibnamefont {Occhialini}}, \bibinfo
  {author} {\bibfnamefont {J.}~\bibnamefont {Pelliciari}}, \bibinfo {author}
  {\bibfnamefont {Y.}~\bibnamefont {Choi}}, \bibinfo {author} {\bibfnamefont
  {T.}~\bibnamefont {Kawaguchi}}, \bibinfo {author} {\bibfnamefont
  {H.}~\bibnamefont {You}}, \bibinfo {author} {\bibfnamefont {J.~F.}\
  \bibnamefont {Mitchell}}, \bibinfo {author} {\bibfnamefont {Y.}~\bibnamefont
  {Shao-Horn}}, \ and\ \bibinfo {author} {\bibfnamefont {R.}~\bibnamefont
  {Comin}},\ }\href {\doibase 10.1103/PhysRevLett.122.017202} {\bibfield
  {journal} {\bibinfo  {journal} {Phys. Rev. Lett.}\ }\textbf {\bibinfo
  {volume} {122}},\ \bibinfo {eid} {017202} (\bibinfo {year}
  {2019})}\BibitemShut {NoStop}%
\bibitem [{\citenamefont {Lovesey}\ \emph {et~al.}(2022)\citenamefont
  {Lovesey}, \citenamefont {Khalyavin},\ and\ \citenamefont {van~der
  Laan}}]{Lovesey2022}%
  \BibitemOpen
  \bibfield  {author} {\bibinfo {author} {\bibfnamefont {S.~W.}\ \bibnamefont
  {Lovesey}}, \bibinfo {author} {\bibfnamefont {D.~D.}\ \bibnamefont
  {Khalyavin}}, \ and\ \bibinfo {author} {\bibfnamefont {G.}~\bibnamefont
  {van~der Laan}},\ }\href {\doibase 10.1103/PhysRevB.105.014403} {\bibfield
  {journal} {\bibinfo  {journal} {Phys. Rev. B}\ }\textbf {\bibinfo {volume}
  {105}},\ \bibinfo {pages} {014403} (\bibinfo {year} {2022})}\BibitemShut
  {NoStop}%
\bibitem [{\citenamefont {Gonz{\'{a}}lez-Hern{\'{a}}ndez}\ \emph
  {et~al.}(2021)\citenamefont {Gonz{\'{a}}lez-Hern{\'{a}}ndez}, \citenamefont
  {{\v{S}}mejkal}, \citenamefont {V{\'{y}}born{\'{y}}}, \citenamefont {Yahagi},
  \citenamefont {Sinova}, \citenamefont {Jungwirth},\ and\ \citenamefont
  {{\v{Z}}elezn{\'{y}}}}]{Gonzalez2021}%
  \BibitemOpen
  \bibfield  {author} {\bibinfo {author} {\bibfnamefont {R.}~\bibnamefont
  {Gonz{\'{a}}lez-Hern{\'{a}}ndez}}, \bibinfo {author} {\bibfnamefont
  {L.}~\bibnamefont {{\v{S}}mejkal}}, \bibinfo {author} {\bibfnamefont
  {K.}~\bibnamefont {V{\'{y}}born{\'{y}}}}, \bibinfo {author} {\bibfnamefont
  {Y.}~\bibnamefont {Yahagi}}, \bibinfo {author} {\bibfnamefont
  {J.}~\bibnamefont {Sinova}}, \bibinfo {author} {\bibfnamefont
  {T.}~\bibnamefont {Jungwirth}}, \ and\ \bibinfo {author} {\bibfnamefont
  {J.}~\bibnamefont {{\v{Z}}elezn{\'{y}}}},\ }\href {\doibase
  10.1103/physrevlett.126.127701} {\bibfield  {journal} {\bibinfo  {journal}
  {Phys. Rev. Lett.}\ }\textbf {\bibinfo {volume} {126}},\ \bibinfo {pages}
  {127701} (\bibinfo {year} {2021})}\BibitemShut {NoStop}%
\bibitem [{\citenamefont {\ifmmode~\check{S}\else \v{S}\fi{}mejkal}\ \emph
  {et~al.}(2022{\natexlab{c}})\citenamefont {\ifmmode~\check{S}\else
  \v{S}\fi{}mejkal}, \citenamefont {Hellenes}, \citenamefont
  {Gonz\'alez-Hern\'andez}, \citenamefont {Sinova},\ and\ \citenamefont
  {Jungwirth}}]{Smejkal2022}%
  \BibitemOpen
  \bibfield  {author} {\bibinfo {author} {\bibfnamefont {L.}~\bibnamefont
  {\ifmmode~\check{S}\else \v{S}\fi{}mejkal}}, \bibinfo {author} {\bibfnamefont
  {A.~B.}\ \bibnamefont {Hellenes}}, \bibinfo {author} {\bibfnamefont
  {R.}~\bibnamefont {Gonz\'alez-Hern\'andez}}, \bibinfo {author} {\bibfnamefont
  {J.}~\bibnamefont {Sinova}}, \ and\ \bibinfo {author} {\bibfnamefont
  {T.}~\bibnamefont {Jungwirth}},\ }\href {\doibase 10.1103/PhysRevX.12.011028}
  {\bibfield  {journal} {\bibinfo  {journal} {Phys. Rev. X}\ }\textbf {\bibinfo
  {volume} {12}},\ \bibinfo {pages} {011028} (\bibinfo {year}
  {2022}{\natexlab{c}})}\BibitemShut {NoStop}%
\bibitem [{\citenamefont {Bose}\ \emph {et~al.}(2022)\citenamefont {Bose},
  \citenamefont {Schreiber}, \citenamefont {Jain}, \citenamefont {Shao},
  \citenamefont {Nair}, \citenamefont {Sun}, \citenamefont {Zhang},
  \citenamefont {Muller}, \citenamefont {Tsymbal}, \citenamefont {Schlom},\
  and\ \citenamefont {Ralph}}]{Bose2022}%
  \BibitemOpen
  \bibfield  {author} {\bibinfo {author} {\bibfnamefont {A.}~\bibnamefont
  {Bose}}, \bibinfo {author} {\bibfnamefont {N.~J.}\ \bibnamefont {Schreiber}},
  \bibinfo {author} {\bibfnamefont {R.}~\bibnamefont {Jain}}, \bibinfo {author}
  {\bibfnamefont {D.-F.}\ \bibnamefont {Shao}}, \bibinfo {author}
  {\bibfnamefont {H.~P.}\ \bibnamefont {Nair}}, \bibinfo {author}
  {\bibfnamefont {J.}~\bibnamefont {Sun}}, \bibinfo {author} {\bibfnamefont
  {X.~S.}\ \bibnamefont {Zhang}}, \bibinfo {author} {\bibfnamefont {D.~A.}\
  \bibnamefont {Muller}}, \bibinfo {author} {\bibfnamefont {E.~Y.}\
  \bibnamefont {Tsymbal}}, \bibinfo {author} {\bibfnamefont {D.~G.}\
  \bibnamefont {Schlom}}, \ and\ \bibinfo {author} {\bibfnamefont {D.~C.}\
  \bibnamefont {Ralph}},\ }\href {\doibase 10.1038/s41928-022-00744-8}
  {\bibfield  {journal} {\bibinfo  {journal} {Nat. Electron.}\ }\textbf
  {\bibinfo {volume} {5}},\ \bibinfo {pages} {267} (\bibinfo {year}
  {2022})}\BibitemShut {NoStop}%
\bibitem [{\citenamefont {Karube}\ \emph {et~al.}(2022)\citenamefont {Karube},
	\citenamefont {Tanaka}, \citenamefont {Sugawara}, \citenamefont {Kadoguchi},
	\citenamefont {Kohda},\ and\ \citenamefont {Nitta}}]{Karube2022}%
\BibitemOpen
\bibfield  {author} {\bibinfo {author} {\bibfnamefont {S.}~\bibnamefont
		{Karube}}, \bibinfo {author} {\bibfnamefont {T.}~\bibnamefont {Tanaka}},
	\bibinfo {author} {\bibfnamefont {D.}~\bibnamefont {Sugawara}}, \bibinfo
	{author} {\bibfnamefont {N.}~\bibnamefont {Kadoguchi}}, \bibinfo {author}
	{\bibfnamefont {M.}~\bibnamefont {Kohda}}, \ and\ \bibinfo {author}
	{\bibfnamefont {J.}~\bibnamefont {Nitta}},\ }\href {\doibase
	10.1103/PhysRevLett.129.137201} {\bibfield  {journal} {\bibinfo  {journal}
		{Phys. Rev. Lett.}\ }\textbf {\bibinfo {volume} {129}},\ \bibinfo {pages}
	{137201} (\bibinfo {year} {2022})}\BibitemShut {NoStop}%
\bibitem [{\citenamefont {Bai}\ \emph {et~al.}(2022)\citenamefont {Bai},
  \citenamefont {Han}, \citenamefont {Feng}, \citenamefont {Zhou},
  \citenamefont {Su}, \citenamefont {Wang}, \citenamefont {Liao}, \citenamefont
  {Zhu}, \citenamefont {Chen}, \citenamefont {Pan}, \citenamefont {Fan},\ and\
  \citenamefont {Song}}]{H-Bai2022}%
  \BibitemOpen
  \bibfield  {author} {\bibinfo {author} {\bibfnamefont {H.}~\bibnamefont
  {Bai}}, \bibinfo {author} {\bibfnamefont {L.}~\bibnamefont {Han}}, \bibinfo
  {author} {\bibfnamefont {X.~Y.}\ \bibnamefont {Feng}}, \bibinfo {author}
  {\bibfnamefont {Y.~J.}\ \bibnamefont {Zhou}}, \bibinfo {author}
  {\bibfnamefont {R.~X.}\ \bibnamefont {Su}}, \bibinfo {author} {\bibfnamefont
  {Q.}~\bibnamefont {Wang}}, \bibinfo {author} {\bibfnamefont {L.~Y.}\
  \bibnamefont {Liao}}, \bibinfo {author} {\bibfnamefont {W.~X.}\ \bibnamefont
  {Zhu}}, \bibinfo {author} {\bibfnamefont {X.~Z.}\ \bibnamefont {Chen}},
  \bibinfo {author} {\bibfnamefont {F.}~\bibnamefont {Pan}}, \bibinfo {author}
  {\bibfnamefont {X.~L.}\ \bibnamefont {Fan}}, \ and\ \bibinfo {author}
  {\bibfnamefont {C.}~\bibnamefont {Song}},\ }\href {\doibase
  10.1103/PhysRevLett.128.197202} {\bibfield  {journal} {\bibinfo  {journal}
  {Phys. Rev. Lett.}\ }\textbf {\bibinfo {volume} {128}},\ \bibinfo {pages}
  {197202} (\bibinfo {year} {2022})}\BibitemShut {NoStop}%
\bibitem [{\citenamefont {Shao}\ \emph {et~al.}(2021)\citenamefont {Shao},
  \citenamefont {Zhang}, \citenamefont {Li}, \citenamefont {Eom},\ and\
  \citenamefont {Tsymbal}}]{DF-Shao2021a}%
  \BibitemOpen
  \bibfield  {author} {\bibinfo {author} {\bibfnamefont {D.-F.}\ \bibnamefont
  {Shao}}, \bibinfo {author} {\bibfnamefont {S.-H.}\ \bibnamefont {Zhang}},
  \bibinfo {author} {\bibfnamefont {M.}~\bibnamefont {Li}}, \bibinfo {author}
  {\bibfnamefont {C.-B.}\ \bibnamefont {Eom}}, \ and\ \bibinfo {author}
  {\bibfnamefont {E.~Y.}\ \bibnamefont {Tsymbal}},\ }\href {\doibase
  10.1038/s41467-021-26915-3} {\bibfield  {journal} {\bibinfo  {journal} {Nat.
  Commun.}\ }\textbf {\bibinfo {volume} {12}},\ \bibinfo {pages} {7061}
  (\bibinfo {year} {2021})}\BibitemShut {NoStop}%
\bibitem [{\citenamefont {Samanta}\ \emph {et~al.}(2020)\citenamefont
  {Samanta}, \citenamefont {Le{\v z}ai{\'c}}, \citenamefont {Merte},
  \citenamefont {Freimuth}, \citenamefont {Bl{\"u}gel},\ and\ \citenamefont
  {Mokrousov}}]{Samanta2020}%
  \BibitemOpen
  \bibfield  {author} {\bibinfo {author} {\bibfnamefont {K.}~\bibnamefont
  {Samanta}}, \bibinfo {author} {\bibfnamefont {M.}~\bibnamefont {Le{\v
  z}ai{\'c}}}, \bibinfo {author} {\bibfnamefont {M.}~\bibnamefont {Merte}},
  \bibinfo {author} {\bibfnamefont {F.}~\bibnamefont {Freimuth}}, \bibinfo
  {author} {\bibfnamefont {S.}~\bibnamefont {Bl{\"u}gel}}, \ and\ \bibinfo
  {author} {\bibfnamefont {Y.}~\bibnamefont {Mokrousov}},\ }\href {\doibase
  10.1063/5.0005017} {\bibfield  {journal} {\bibinfo  {journal} {J. Appl.
  Phys.}\ }\textbf {\bibinfo {volume} {127}},\ \bibinfo {pages} {213904}
  (\bibinfo {year} {2020})}\BibitemShut {NoStop}%
\bibitem [{\citenamefont {Zhou}\ \emph {et~al.}(2021)\citenamefont {Zhou},
  \citenamefont {Feng}, \citenamefont {Yang}, \citenamefont {Guo},\ and\
  \citenamefont {Yao}}]{XD-Zhou2021}%
  \BibitemOpen
  \bibfield  {author} {\bibinfo {author} {\bibfnamefont {X.}~\bibnamefont
  {Zhou}}, \bibinfo {author} {\bibfnamefont {W.}~\bibnamefont {Feng}}, \bibinfo
  {author} {\bibfnamefont {X.}~\bibnamefont {Yang}}, \bibinfo {author}
  {\bibfnamefont {G.-Y.}\ \bibnamefont {Guo}}, \ and\ \bibinfo {author}
  {\bibfnamefont {Y.}~\bibnamefont {Yao}},\ }\href {\doibase
  10.1103/PhysRevB.104.024401} {\bibfield  {journal} {\bibinfo  {journal}
  {Phys. Rev. B.}\ }\textbf {\bibinfo {volume} {104}},\ \bibinfo {pages}
  {024401} (\bibinfo {year} {2021})}\BibitemShut {NoStop}%
\bibitem [{\citenamefont {Shao}\ \emph {et~al.}(2022)\citenamefont {Shao},
  \citenamefont {Zhang}, \citenamefont {Xiao}, \citenamefont {Wang},
  \citenamefont {Lu}, \citenamefont {Sun},\ and\ \citenamefont
  {Tsymbal}}]{DF-Shao2022}%
  \BibitemOpen
  \bibfield  {author} {\bibinfo {author} {\bibfnamefont {D.-F.}\ \bibnamefont
  {Shao}}, \bibinfo {author} {\bibfnamefont {S.-H.}\ \bibnamefont {Zhang}},
  \bibinfo {author} {\bibfnamefont {R.-C.}\ \bibnamefont {Xiao}}, \bibinfo
  {author} {\bibfnamefont {Z.-A.}\ \bibnamefont {Wang}}, \bibinfo {author}
  {\bibfnamefont {W.~J.}\ \bibnamefont {Lu}}, \bibinfo {author} {\bibfnamefont
  {Y.~P.}\ \bibnamefont {Sun}}, \ and\ \bibinfo {author} {\bibfnamefont
  {E.~Y.}\ \bibnamefont {Tsymbal}},\ }\href {\doibase
  10.1103/PhysRevB.106.L180404} {\bibfield  {journal} {\bibinfo  {journal}
  {Phys. Rev. B}\ }\textbf {\bibinfo {volume} {106}},\ \bibinfo {pages}
  {L180404} (\bibinfo {year} {2022})}\BibitemShut {NoStop}%
\bibitem [{\citenamefont {Wadley}\ \emph {et~al.}(2016)\citenamefont {Wadley},
  \citenamefont {Howells}, \citenamefont {{\v Z}elezn{\'y}}, \citenamefont
  {Andrews}, \citenamefont {Hills}, \citenamefont {Campion}, \citenamefont
  {Nov{\'a}k}, \citenamefont {Olejn{\'{\i}}k}, \citenamefont {Maccherozzi},
  \citenamefont {Dhesi}, \citenamefont {Martin}, \citenamefont {Wagner},
  \citenamefont {Wunderlich}, \citenamefont {Freimuth}, \citenamefont
  {Mokrousov}, \citenamefont {Kune{\v s}}, \citenamefont {Chauhan},
  \citenamefont {Grzybowski}, \citenamefont {Rushforth}, \citenamefont
  {Edmonds}, \citenamefont {Gallagher},\ and\ \citenamefont
  {Jungwirth}}]{Wadley2016}%
  \BibitemOpen
  \bibfield  {author} {\bibinfo {author} {\bibfnamefont {P.}~\bibnamefont
  {Wadley}}, \bibinfo {author} {\bibfnamefont {B.}~\bibnamefont {Howells}},
  \bibinfo {author} {\bibfnamefont {J.}~\bibnamefont {{\v Z}elezn{\'y}}},
  \bibinfo {author} {\bibfnamefont {C.}~\bibnamefont {Andrews}}, \bibinfo
  {author} {\bibfnamefont {V.}~\bibnamefont {Hills}}, \bibinfo {author}
  {\bibfnamefont {R.~P.}\ \bibnamefont {Campion}}, \bibinfo {author}
  {\bibfnamefont {V.}~\bibnamefont {Nov{\'a}k}}, \bibinfo {author}
  {\bibfnamefont {K.}~\bibnamefont {Olejn{\'{\i}}k}}, \bibinfo {author}
  {\bibfnamefont {F.}~\bibnamefont {Maccherozzi}}, \bibinfo {author}
  {\bibfnamefont {S.~S.}\ \bibnamefont {Dhesi}}, \bibinfo {author}
  {\bibfnamefont {S.~Y.}\ \bibnamefont {Martin}}, \bibinfo {author}
  {\bibfnamefont {T.}~\bibnamefont {Wagner}}, \bibinfo {author} {\bibfnamefont
  {J.}~\bibnamefont {Wunderlich}}, \bibinfo {author} {\bibfnamefont
  {F.}~\bibnamefont {Freimuth}}, \bibinfo {author} {\bibfnamefont
  {Y.}~\bibnamefont {Mokrousov}}, \bibinfo {author} {\bibfnamefont
  {J.}~\bibnamefont {Kune{\v s}}}, \bibinfo {author} {\bibfnamefont {J.~S.}\
  \bibnamefont {Chauhan}}, \bibinfo {author} {\bibfnamefont {M.~J.}\
  \bibnamefont {Grzybowski}}, \bibinfo {author} {\bibfnamefont {A.~W.}\
  \bibnamefont {Rushforth}}, \bibinfo {author} {\bibfnamefont {K.~W.}\
  \bibnamefont {Edmonds}}, \bibinfo {author} {\bibfnamefont {B.~L.}\
  \bibnamefont {Gallagher}}, \ and\ \bibinfo {author} {\bibfnamefont
  {T.}~\bibnamefont {Jungwirth}},\ }\href {\doibase 10.1126/science.aab1031}
  {\bibfield  {journal} {\bibinfo  {journal} {Science}\ }\textbf {\bibinfo
  {volume} {351}},\ \bibinfo {pages} {587} (\bibinfo {year}
  {2016})}\BibitemShut {NoStop}%
\bibitem [{\citenamefont {Godinho}\ \emph {et~al.}(2018)\citenamefont
  {Godinho}, \citenamefont {Reichlov{\'a}}, \citenamefont {Kriegner},
  \citenamefont {Nov{\'a}k}, \citenamefont {Olejn{\'{\i}}k}, \citenamefont
  {Ka{\v s}par}, \citenamefont {{\v S}ob{\'a}{\v n}}, \citenamefont {Wadley},
  \citenamefont {Campion}, \citenamefont {Otxoa}, \citenamefont {Roy},
  \citenamefont {{\v Z}elezn{\'y}}, \citenamefont {Jungwirth},\ and\
  \citenamefont {Wunderlich}}]{Godinho2018}%
  \BibitemOpen
  \bibfield  {author} {\bibinfo {author} {\bibfnamefont {J.}~\bibnamefont
  {Godinho}}, \bibinfo {author} {\bibfnamefont {H.}~\bibnamefont
  {Reichlov{\'a}}}, \bibinfo {author} {\bibfnamefont {D.}~\bibnamefont
  {Kriegner}}, \bibinfo {author} {\bibfnamefont {V.}~\bibnamefont {Nov{\'a}k}},
  \bibinfo {author} {\bibfnamefont {K.}~\bibnamefont {Olejn{\'{\i}}k}},
  \bibinfo {author} {\bibfnamefont {Z.}~\bibnamefont {Ka{\v s}par}}, \bibinfo
  {author} {\bibfnamefont {Z.}~\bibnamefont {{\v S}ob{\'a}{\v n}}}, \bibinfo
  {author} {\bibfnamefont {P.}~\bibnamefont {Wadley}}, \bibinfo {author}
  {\bibfnamefont {R.~P.}\ \bibnamefont {Campion}}, \bibinfo {author}
  {\bibfnamefont {R.~M.}\ \bibnamefont {Otxoa}}, \bibinfo {author}
  {\bibfnamefont {P.~E.}\ \bibnamefont {Roy}}, \bibinfo {author} {\bibfnamefont
  {J.}~\bibnamefont {{\v Z}elezn{\'y}}}, \bibinfo {author} {\bibfnamefont
  {T.}~\bibnamefont {Jungwirth}}, \ and\ \bibinfo {author} {\bibfnamefont
  {J.}~\bibnamefont {Wunderlich}},\ }\href {\doibase
  10.1038/s41467-018-07092-2} {\bibfield  {journal} {\bibinfo  {journal} {Nat.
  Commun.}\ }\textbf {\bibinfo {volume} {9}},\ \bibinfo {eid} {4686} (\bibinfo
  {year} {2018})}\BibitemShut {NoStop}%
\bibitem [{\citenamefont {Chen}\ \emph {et~al.}(2019)\citenamefont {Chen},
  \citenamefont {Zhou}, \citenamefont {Cheng}, \citenamefont {Song},
  \citenamefont {Zhang}, \citenamefont {Wu}, \citenamefont {Ba}, \citenamefont
  {Li}, \citenamefont {Sun}, \citenamefont {You}, \citenamefont {Zhao},\ and\
  \citenamefont {Pan}}]{XZ-Chen2019}%
  \BibitemOpen
  \bibfield  {author} {\bibinfo {author} {\bibfnamefont {X.}~\bibnamefont
  {Chen}}, \bibinfo {author} {\bibfnamefont {X.}~\bibnamefont {Zhou}}, \bibinfo
  {author} {\bibfnamefont {R.}~\bibnamefont {Cheng}}, \bibinfo {author}
  {\bibfnamefont {C.}~\bibnamefont {Song}}, \bibinfo {author} {\bibfnamefont
  {J.}~\bibnamefont {Zhang}}, \bibinfo {author} {\bibfnamefont
  {Y.}~\bibnamefont {Wu}}, \bibinfo {author} {\bibfnamefont {Y.}~\bibnamefont
  {Ba}}, \bibinfo {author} {\bibfnamefont {H.}~\bibnamefont {Li}}, \bibinfo
  {author} {\bibfnamefont {Y.}~\bibnamefont {Sun}}, \bibinfo {author}
  {\bibfnamefont {Y.}~\bibnamefont {You}}, \bibinfo {author} {\bibfnamefont
  {Y.}~\bibnamefont {Zhao}}, \ and\ \bibinfo {author} {\bibfnamefont
  {F.}~\bibnamefont {Pan}},\ }\href {\doibase 10.1038/s41563-019-0424-2}
  {\bibfield  {journal} {\bibinfo  {journal} {Nat. Mater.}\ }\textbf {\bibinfo
  {volume} {18}},\ \bibinfo {pages} {931} (\bibinfo {year} {2019})}\BibitemShut
  {NoStop}%
\bibitem [{\citenamefont {Ashcroft}\ and\ \citenamefont
  {Mermin}(1976)}]{Ashcroft1976}%
  \BibitemOpen
  \bibfield  {author} {\bibinfo {author} {\bibfnamefont {N.~W.}\ \bibnamefont
  {Ashcroft}}\ and\ \bibinfo {author} {\bibfnamefont {N.~D.}\ \bibnamefont
  {Mermin}},\ }\href@noop {} {\emph {\bibinfo {title} {Solid State Physics}}}\
  (\bibinfo  {publisher} {Saunders College Publishing, Philadelphia},\ \bibinfo
  {year} {1976})\BibitemShut {NoStop}%
\bibitem [{\citenamefont {van Houten}\ \emph {et~al.}(1992)\citenamefont {van
  Houten}, \citenamefont {Molenkamp}, \citenamefont {Beenakker},\ and\
  \citenamefont {Foxon}}]{Houten1992}%
  \BibitemOpen
  \bibfield  {author} {\bibinfo {author} {\bibfnamefont {H.}~\bibnamefont {van
  Houten}}, \bibinfo {author} {\bibfnamefont {L.~W.}\ \bibnamefont
  {Molenkamp}}, \bibinfo {author} {\bibfnamefont {C.~W.~J.}\ \bibnamefont
  {Beenakker}}, \ and\ \bibinfo {author} {\bibfnamefont {C.~T.}\ \bibnamefont
  {Foxon}},\ }\href {\doibase 10.1088/0268-1242/7/3b/052} {\bibfield  {journal}
  {\bibinfo  {journal} {Semicond. Sci. Technol.}\ }\textbf {\bibinfo {volume}
  {7}},\ \bibinfo {pages} {B215} (\bibinfo {year} {1992})}\BibitemShut
  {NoStop}%
\bibitem [{\citenamefont {Behnia}(2015)}]{Behnia2015}%
  \BibitemOpen
  \bibfield  {author} {\bibinfo {author} {\bibfnamefont {K.}~\bibnamefont
  {Behnia}},\ }\href@noop {} {\emph {\bibinfo {title} {Fundamentals of
  Thermoelectricity}}}\ (\bibinfo  {publisher} {Oxford University Press,
  Oxford},\ \bibinfo {year} {2015})\BibitemShut {NoStop}%
\bibitem [{Sup()}]{SuppMater}%
  \BibitemOpen
  \href@noop {} {\bibinfo  {journal} {See Supplemental Material at
  http://link.aps.org/xxx, which includes a detailed description of
  computational methods, supplemental figures, and Refs.~\cite{Berlijn2017,Bloechl1994,Kresse1996a,Perdew1996,Dudarev1998,Mostofi2008,YG-Yao2004}}\ }\BibitemShut {NoStop}%
\bibitem [{\citenamefont {Bl{\"o}chl}(1994)}]{Bloechl1994}%
\BibitemOpen
\bibfield  {author} {\bibinfo {author} {\bibfnamefont {P.~E.}\ \bibnamefont
		{Bl{\"o}chl}},\ }\href {http://dx.doi.org/10.1103/physrevb.50.17953}
{\bibfield  {journal} {\bibinfo  {journal} {Phys. Rev. B}\ }\textbf {\bibinfo
		{volume} {50}},\ \bibinfo {pages} {17953} (\bibinfo {year}
	{1994})}\BibitemShut {NoStop}%
\bibitem [{\citenamefont {Kresse}\ and\ \citenamefont
	{Furthm{\"u}ller}(1996)}]{Kresse1996a}%
\BibitemOpen
\bibfield  {author} {\bibinfo {author} {\bibfnamefont {G.}~\bibnamefont
		{Kresse}}\ and\ \bibinfo {author} {\bibfnamefont {J.}~\bibnamefont
		{Furthm{\"u}ller}},\ }\href {\doibase 10.1016/0927-0256(96)00008-0}
{\bibfield  {journal} {\bibinfo  {journal} {Comput. Mater. Sci.}\ }\textbf
	{\bibinfo {volume} {6}},\ \bibinfo {pages} {15} (\bibinfo {year}
	{1996})}\BibitemShut {NoStop}%
\bibitem [{\citenamefont {Perdew}\ \emph {et~al.}(1996)\citenamefont {Perdew},
	\citenamefont {Burke},\ and\ \citenamefont {Ernzerhof}}]{Perdew1996}%
\BibitemOpen
\bibfield  {author} {\bibinfo {author} {\bibfnamefont {J.~P.}\ \bibnamefont
		{Perdew}}, \bibinfo {author} {\bibfnamefont {K.}~\bibnamefont {Burke}}, \
	and\ \bibinfo {author} {\bibfnamefont {M.}~\bibnamefont {Ernzerhof}},\ }\href
{http://dx.doi.org/10.1103/physrevlett.77.3865} {\bibfield  {journal}
	{\bibinfo  {journal} {Phys. Rev. Lett.}\ }\textbf {\bibinfo {volume} {77}},\
	\bibinfo {pages} {3865} (\bibinfo {year} {1996})}\BibitemShut {NoStop}%
\bibitem [{\citenamefont {Dudarev}\ \emph {et~al.}(1998)\citenamefont
	{Dudarev}, \citenamefont {Botton}, \citenamefont {Savrasov}, \citenamefont
	{Humphreys},\ and\ \citenamefont {Sutton}}]{Dudarev1998}%
\BibitemOpen
\bibfield  {author} {\bibinfo {author} {\bibfnamefont {S.~L.}\ \bibnamefont
		{Dudarev}}, \bibinfo {author} {\bibfnamefont {G.~A.}\ \bibnamefont {Botton}},
	\bibinfo {author} {\bibfnamefont {S.~Y.}\ \bibnamefont {Savrasov}}, \bibinfo
	{author} {\bibfnamefont {C.~J.}\ \bibnamefont {Humphreys}}, \ and\ \bibinfo
	{author} {\bibfnamefont {A.~P.}\ \bibnamefont {Sutton}},\ }\href {\doibase
	10.1103/PhysRevB.57.1505} {\bibfield  {journal} {\bibinfo  {journal} {Phys.
			Rev. B}\ }\textbf {\bibinfo {volume} {57}},\ \bibinfo {pages} {1505}
	(\bibinfo {year} {1998})}\BibitemShut {NoStop}%
\bibitem [{\citenamefont {Mostofi}\ \emph {et~al.}(2008)\citenamefont
	{Mostofi}, \citenamefont {Yates}, \citenamefont {Lee}, \citenamefont {Souza},
	\citenamefont {Vanderbilt},\ and\ \citenamefont {Marzari}}]{Mostofi2008}%
\BibitemOpen
\bibfield  {author} {\bibinfo {author} {\bibfnamefont {A.~A.}\ \bibnamefont
		{Mostofi}}, \bibinfo {author} {\bibfnamefont {J.~R.}\ \bibnamefont {Yates}},
	\bibinfo {author} {\bibfnamefont {Y.-S.}\ \bibnamefont {Lee}}, \bibinfo
	{author} {\bibfnamefont {I.}~\bibnamefont {Souza}}, \bibinfo {author}
	{\bibfnamefont {D.}~\bibnamefont {Vanderbilt}}, \ and\ \bibinfo {author}
	{\bibfnamefont {N.}~\bibnamefont {Marzari}},\ }\href
{http://dx.doi.org/10.1016/j.cpc.2007.11.016} {\bibfield  {journal} {\bibinfo
		{journal} {Comput. Phys. Commun.}\ }\textbf {\bibinfo {volume} {178}},\
	\bibinfo {pages} {685} (\bibinfo {year} {2008})}\BibitemShut {NoStop}%
\bibitem [{\citenamefont {Yao}\ \emph {et~al.}(2004)\citenamefont {Yao},
	\citenamefont {Kleinman}, \citenamefont {MacDonald}, \citenamefont {Sinova},
	\citenamefont {Jungwirth}, \citenamefont {Wang}, \citenamefont {Wang},\ and\
	\citenamefont {Niu}}]{YG-Yao2004}%
\BibitemOpen
\bibfield  {author} {\bibinfo {author} {\bibfnamefont {Y.}~\bibnamefont
		{Yao}}, \bibinfo {author} {\bibfnamefont {L.}~\bibnamefont {Kleinman}},
	\bibinfo {author} {\bibfnamefont {A.~H.}\ \bibnamefont {MacDonald}}, \bibinfo
	{author} {\bibfnamefont {J.}~\bibnamefont {Sinova}}, \bibinfo {author}
	{\bibfnamefont {T.}~\bibnamefont {Jungwirth}}, \bibinfo {author}
	{\bibfnamefont {D.-S.}\ \bibnamefont {Wang}}, \bibinfo {author}
	{\bibfnamefont {E.}~\bibnamefont {Wang}}, \ and\ \bibinfo {author}
	{\bibfnamefont {Q.}~\bibnamefont {Niu}},\ }\href
{https://doi.org/10.1103/PhysRevLett.92.037204} {\bibfield  {journal}
	{\bibinfo  {journal} {Phys. Rev. Lett.}\ }\textbf {\bibinfo {volume} {92}},\
	\bibinfo {pages} {037204} (\bibinfo {year} {2004})}\BibitemShut {NoStop}%
\bibitem [{\citenamefont {Ahn}\ \emph {et~al.}(2019)\citenamefont {Ahn},
  \citenamefont {Hariki}, \citenamefont {Lee},\ and\ \citenamefont
  {Kune\ifmmode~\check{s}\else \v{s}\fi{}}}]{Ahn2019}%
  \BibitemOpen
\bibfield  {journal} {  }\bibfield  {author} {\bibinfo {author} {\bibfnamefont
  {K.-H.}\ \bibnamefont {Ahn}}, \bibinfo {author} {\bibfnamefont
  {A.}~\bibnamefont {Hariki}}, \bibinfo {author} {\bibfnamefont {K.-W.}\
  \bibnamefont {Lee}}, \ and\ \bibinfo {author} {\bibfnamefont
  {J.}~\bibnamefont {Kune\ifmmode~\check{s}\else \v{s}\fi{}}},\ }\href
  {\doibase 10.1103/PhysRevB.99.184432} {\bibfield  {journal} {\bibinfo
  {journal} {Phys. Rev. B}\ }\textbf {\bibinfo {volume} {99}},\ \bibinfo
  {pages} {184432} (\bibinfo {year} {2019})}\BibitemShut {NoStop}%
\bibitem [{\citenamefont {Zhan}\ \emph {et~al.}(2023)\citenamefont {Zhan},
	\citenamefont {Li}, \citenamefont {Shi}, \citenamefont {Chen},\ and\
	\citenamefont {Sun}}]{J-Zhan2023}%
\BibitemOpen
\bibfield  {author} {\bibinfo {author} {\bibfnamefont {J.}~\bibnamefont
		{Zhan}}, \bibinfo {author} {\bibfnamefont {J.}~\bibnamefont {Li}}, \bibinfo
	{author} {\bibfnamefont {W.}~\bibnamefont {Shi}}, \bibinfo {author}
	{\bibfnamefont {X.-Q.}\ \bibnamefont {Chen}}, \ and\ \bibinfo {author}
	{\bibfnamefont {Y.}~\bibnamefont {Sun}},\ }\href {\doibase
	10.1103/PhysRevB.107.224402} {\bibfield  {journal} {\bibinfo  {journal}
		{Phys. Rev. B}\ }\textbf {\bibinfo {volume} {107}},\ \bibinfo {pages}
	{224402} (\bibinfo {year} {2023})}\BibitemShut {NoStop}%
\bibitem [{\citenamefont {Mazin}\ \emph {et~al.}(2021)\citenamefont {Mazin},
  \citenamefont {Koepernik}, \citenamefont {Johannes}, \citenamefont
  {Gonz{\'{a}}lez-Hern{\'{a}}ndez},\ and\ \citenamefont
  {{\v{S}}mejkal}}]{Mazin2021}%
  \BibitemOpen
\bibfield  {journal} {  }\bibfield  {author} {\bibinfo {author} {\bibfnamefont
  {I.~I.}\ \bibnamefont {Mazin}}, \bibinfo {author} {\bibfnamefont
  {K.}~\bibnamefont {Koepernik}}, \bibinfo {author} {\bibfnamefont {M.~D.}\
  \bibnamefont {Johannes}}, \bibinfo {author} {\bibfnamefont {R.}~\bibnamefont
  {Gonz{\'{a}}lez-Hern{\'{a}}ndez}}, \ and\ \bibinfo {author} {\bibfnamefont
  {L.}~\bibnamefont {{\v{S}}mejkal}},\ }\href {\doibase
  10.1073/pnas.2108924118} {\bibfield  {journal} {\bibinfo  {journal} {Proc.
  Natl. Acad. Sci. (USA)}\ }\textbf {\bibinfo {volume} {118}},\ \bibinfo
  {pages} {e2108924118} (\bibinfo {year} {2021})}\BibitemShut {NoStop}%
\bibitem [{\citenamefont {Zhang}\ \emph
  {et~al.}(2021{\natexlab{b}})\citenamefont {Zhang}, \citenamefont {Ma},
  \citenamefont {Zhang},\ and\ \citenamefont {Yao}}]{RW-Zhang2021}%
  \BibitemOpen
  \bibfield  {author} {\bibinfo {author} {\bibfnamefont {R.-W.}\ \bibnamefont
  {Zhang}}, \bibinfo {author} {\bibfnamefont {D.-S.}\ \bibnamefont {Ma}},
  \bibinfo {author} {\bibfnamefont {J.-M.}\ \bibnamefont {Zhang}}, \ and\
  \bibinfo {author} {\bibfnamefont {Y.}~\bibnamefont {Yao}},\ }\href {\doibase
  10.1103/PhysRevB.103.195115} {\bibfield  {journal} {\bibinfo  {journal}
  {Phys. Rev. B}\ }\textbf {\bibinfo {volume} {103}},\ \bibinfo {pages}
  {195115} (\bibinfo {year} {2021}{\natexlab{b}})}\BibitemShut {NoStop}%
\bibitem [{not({\natexlab{a}})}]{note1}%
  \BibitemOpen
  \href@noop {} {\bibfield  {journal} {\bibinfo  {journal} {The discussed spin
  splitting is rather weak, and it is slightly further enhanced by SOC. This
  situations is in sharp contrast to strong spin-split states close to the
  pseudo-nodal surfaces.}\ } ({\natexlab{a}})}\BibitemShut {NoStop}%
\bibitem [{\citenamefont {Zhang}\ \emph {et~al.}(2011)\citenamefont {Zhang},
  \citenamefont {Freimuth}, \citenamefont {Bl\"ugel}, \citenamefont
  {Mokrousov},\ and\ \citenamefont {Souza}}]{HB-Zhang2011}%
  \BibitemOpen
  \bibfield  {author} {\bibinfo {author} {\bibfnamefont {H.}~\bibnamefont
  {Zhang}}, \bibinfo {author} {\bibfnamefont {F.}~\bibnamefont {Freimuth}},
  \bibinfo {author} {\bibfnamefont {S.}~\bibnamefont {Bl\"ugel}}, \bibinfo
  {author} {\bibfnamefont {Y.}~\bibnamefont {Mokrousov}}, \ and\ \bibinfo
  {author} {\bibfnamefont {I.}~\bibnamefont {Souza}},\ }\href {\doibase
  10.1103/PhysRevLett.106.117202} {\bibfield  {journal} {\bibinfo  {journal}
  {Phys. Rev. Lett.}\ }\textbf {\bibinfo {volume} {106}},\ \bibinfo {pages}
  {117202} (\bibinfo {year} {2011})}\BibitemShut {NoStop}%
\bibitem [{not({\natexlab{b}})}]{note2}%
  \BibitemOpen
  \href@noop {} {\bibfield  {journal} {\bibinfo  {journal} {The anomalous
  Nernst and anomalous thermal Hall conductivities in collinear altermagnetic
  RuO$_2$ are larger than the corresponding values in intensively studied
  noncollinear antiferromagnetic Mn$_3$Sn (about 0.2 AK$^{-1}$m$^{-1}$ and 4
  $\times 10^{-2}$ WK$^{-1}$m$^{-1}$, respectively~\cite{XK-Li2017}) and
  Mn$_3$Ge (about 0.3 AK$^{-1}$m$^{-1}$ and 1.5 $\times 10^{-2}$
  WK$^{-1}$m$^{-1}$, respectively~\cite{LC-Xu2020})}\ }
  ({\natexlab{b}})}\BibitemShut {NoStop}%
\bibitem [{\citenamefont {Strohm}\ \emph {et~al.}(2005)\citenamefont {Strohm},
  \citenamefont {Rikken},\ and\ \citenamefont {Wyder}}]{Strohm2005}%
  \BibitemOpen
  \bibfield  {author} {\bibinfo {author} {\bibfnamefont {C.}~\bibnamefont
  {Strohm}}, \bibinfo {author} {\bibfnamefont {G.}~\bibnamefont {Rikken}}, \
  and\ \bibinfo {author} {\bibfnamefont {P.}~\bibnamefont {Wyder}},\ }\href
  {https://doi.org/10.1103/PhysRevLett.95.155901} {\bibfield  {journal}
  {\bibinfo  {journal} {Phys. Rev. Lett}\ }\textbf {\bibinfo {volume} {95}},\
  \bibinfo {pages} {155901} (\bibinfo {year} {2005})}\BibitemShut {NoStop}%
\bibitem [{\citenamefont {Saito}\ \emph {et~al.}(2019)\citenamefont {Saito},
  \citenamefont {Misaki}, \citenamefont {Ishizuka},\ and\ \citenamefont
  {Nagaosa}}]{Saito2019}%
  \BibitemOpen
  \bibfield  {author} {\bibinfo {author} {\bibfnamefont {T.}~\bibnamefont
  {Saito}}, \bibinfo {author} {\bibfnamefont {K.}~\bibnamefont {Misaki}},
  \bibinfo {author} {\bibfnamefont {H.}~\bibnamefont {Ishizuka}}, \ and\
  \bibinfo {author} {\bibfnamefont {N.}~\bibnamefont {Nagaosa}},\ }\href
  {https://doi.org/10.1103/PhysRevLett.123.255901} {\bibfield  {journal}
  {\bibinfo  {journal} {Phys. Rev. Lett}\ }\textbf {\bibinfo {volume} {123}},\
  \bibinfo {pages} {255901} (\bibinfo {year} {2019})}\BibitemShut {NoStop}%
\bibitem [{\citenamefont {Onose}\ \emph {et~al.}(2010)\citenamefont {Onose},
  \citenamefont {Ideue}, \citenamefont {Katsura}, \citenamefont {Shiomi},
  \citenamefont {Nagaosa},\ and\ \citenamefont {Tokura}}]{Onose2010}%
  \BibitemOpen
  \bibfield  {author} {\bibinfo {author} {\bibfnamefont {Y.}~\bibnamefont
  {Onose}}, \bibinfo {author} {\bibfnamefont {T.}~\bibnamefont {Ideue}},
  \bibinfo {author} {\bibfnamefont {H.}~\bibnamefont {Katsura}}, \bibinfo
  {author} {\bibfnamefont {Y.}~\bibnamefont {Shiomi}}, \bibinfo {author}
  {\bibfnamefont {N.}~\bibnamefont {Nagaosa}}, \ and\ \bibinfo {author}
  {\bibfnamefont {Y.}~\bibnamefont {Tokura}},\ }\href
  {https://doi.org/10.1126/science.1188260} {\bibfield  {journal} {\bibinfo
  {journal} {Science}\ }\textbf {\bibinfo {volume} {329}},\ \bibinfo {pages}
  {297} (\bibinfo {year} {2010})}\BibitemShut {NoStop}%
\bibitem [{\citenamefont {Zhang}\ \emph {et~al.}(2019)\citenamefont {Zhang},
  \citenamefont {Zhang}, \citenamefont {Okamoto},\ and\ \citenamefont
  {Xiao}}]{X-Zhang2019}%
  \BibitemOpen
  \bibfield  {author} {\bibinfo {author} {\bibfnamefont {X.}~\bibnamefont
  {Zhang}}, \bibinfo {author} {\bibfnamefont {Y.}~\bibnamefont {Zhang}},
  \bibinfo {author} {\bibfnamefont {S.}~\bibnamefont {Okamoto}}, \ and\
  \bibinfo {author} {\bibfnamefont {D.}~\bibnamefont {Xiao}},\ }\href
  {https://doi.org/10.1103/PhysRevLett.123.167202} {\bibfield  {journal}
  {\bibinfo  {journal} {Phys. Rev. Lett.}\ }\textbf {\bibinfo {volume} {123}},\
  \bibinfo {pages} {167202} (\bibinfo {year} {2019})}\BibitemShut {NoStop}%
\bibitem [{\citenamefont {Krempask{\`y}~et al}()}]{Krempasky2023}%
  \BibitemOpen
  \bibfield  {author} {\bibinfo {author} {\bibfnamefont {J.}~\bibnamefont
  {Krempask{\`y}~et al}},\ }\href {https://doi.org/10.48550/arXiv.2308.10681}
  {\bibinfo  {journal} {arXiv:2308.10681}\ }\BibitemShut {NoStop}%  
\end{thebibliography}


\end{document}